\documentclass[11pt]{article}

\usepackage[letterpaper,margin=1in]{geometry}
\usepackage{setspace}
\usepackage[sort,compress]{cite}

\usepackage{color}
\definecolor{myurlcolor}{rgb}{0.4,0,0}
\definecolor{mycitecolor}{rgb}{0,0.4,0}
\definecolor{mylinkcolor}{rgb}{0,0,0.4}

\usepackage[colorlinks=true, urlcolor = myurlcolor, citecolor = mycitecolor, linkcolor= mylinkcolor, bookmarks=false]{hyperref}
\usepackage{amsfonts,amsmath,amssymb}
\usepackage{graphicx}
\usepackage{dsfont}

\usepackage{arydshln}

\usepackage{xspace}

\usepackage{relsize}
\usepackage{braket}

\usepackage{bm}

\usepackage[bottom]{footmisc}

\usepackage{stmaryrd}

\usepackage{nicefrac}

\newcommand{\email}[1]{\href{mailto:#1}{#1}}

\newcommand{\namedref}[2]{\hyperref[#2]{#1~\ref*{#2}}}
\newcommand{\secref}[1]{\namedref{Sec.}{#1}}
\newcommand{\appref}[1]{\namedref{App.}{#1}}

\newcommand{\figref}[1]{\namedref{Fig.}{#1}}

\begin{document}

\thispagestyle{empty}

\vfill

\begin{center}
{\Large \textbf{%
The exceptional origin of the strange metal\\\smallskip 
and the LFL--HFL transition}}

\vspace{2em}

\renewcommand*{\thefootnote}{\fnsymbol{footnote}}
{\large Eoin Quinn\footnote{Electronic address: \email{eoin.quinn.physics@gmail.com}}}
\renewcommand*{\thefootnote}{\arabic{footnote}}\setcounter{footnote}{0}

\vspace{0.5em}

{\it Independent researcher, London, UK}

\vspace{0.8em}

\today

\end{center}

\vspace{2.5em}

\begin{center}
\textbf{Abstract}\\

\vspace{1em}

\begin{minipage}{0.8\textwidth}
We propose an algebraic framework for the strange metal regime of strongly correlated electrons. We show that the exceptional superconformal algebra $D(2,1;\alpha)$ admits two distinct contractions of its conformal sector: one to a pair of canonical fermions, the underlying degrees of freedom of the Landau--Fermi liquid (LFL), and one to the algebra of Hubbard operators, which characterise a distinct metallic regime, the Hubbard--Fermi liquid (HFL). We argue that competition between these two metallic states drives the emergence of the strange metal as a $0+1$D superconformal bath. We analyse the resulting thermodynamics, and at leading order obtain a parameter-free prediction, $4\pi^2\gamma^{-1} =\chi_s^{-1} + \chi_c^{-1}$, relating the Sommerfeld coefficient to the static spin and charge susceptibilities. We further show that the LFL--HFL transition is discontinuous at low temperature, forced by an algebraic constraint on the Hubbard contraction, and map out the resulting phase diagram. We connect the framework to microscopic lattice models and to the phenomenology of correlated insulators.  
\end{minipage}
\end{center}

\vfill

\newpage
\thispagestyle{empty}
{\setstretch{0.98}
\tableofcontents
}
\newpage
\setcounter{page}{1}

\section{Introduction}
\label{sec:intro}

The quantum theory of metals rests upon an algebraic assumption: that the electronic correlations are governed by the canonical anti-commutation relations, $\{c_\sigma, c^\dagger_{\sigma'}\} = \delta_{\sigma\sigma'}$. This underpins the microscopic derivation of Landau--Fermi liquid (LFL) theory, and its reduction of the quantum many-body problem to coherent quasiparticles. Yet in strongly correlated systems, such as the cuprates and heavy fermion compounds, this framework systematically fails \cite{Keimer2015, Stewart2001}. A prominent example is the strange metal, a macroscopic regime whose anomalous thermodynamics have resisted a canonical description \cite{Phillips2022}.

We argue that this failure can be traced algebraically. The electron itself admits a second distinct algebraic formulation, introduced by Hubbard in 1965 \cite{Hubbard1965}. In previous work we cast this as the Lie superalgebra $\mathfrak{su}(2|2)$, and proposed that it governs a metallic state distinct from the LFL, termed the Hubbard--Fermi liquid (HFL) \cite{Quinn2018, Quinn2021}. The Hubbard interaction enters as an emergent chemical potential, splitting the electronic band in two. A consequence is that the electronic Luttinger sum rule, which pins the volume enclosed by the LFL's Fermi surface, is intrinsically violated here. The two metallic states are not adiabatically connected, and this LFL--HFL transition is the central focus of this work.

We propose an algebraic framework that connects the canonical and Hubbard formulations of the electron. We show that both descend from the exceptional Lie superalgebra $D(2,1;\alpha)$, as distinct contractions of its conformal sector. This algebra is distinguished: it is the only simple Lie superalgebra with continuously deformable structure constants \cite{Kac1977}, parameterised by $\alpha$, and we find that this deformability governs the LFL--HFL transition. The algebra's conformal sector encodes a purely temporal scale invariance, which we conjecture arises dynamically from the competing coherence scales of the two metallic states, yielding an emergent $0+1$D superconformal bath. Evaluating the resulting Schwarzian thermodynamics in the semiclassical regime, we derive a parameter-free thermodynamic relation:
\begin{equation}\label{eq:ETR_intro}
\frac{4\pi^2 k_B^2}{\gamma} = \frac{(g\mu_B)^2}{\chi_s} + \frac{4e^2}{\chi_c} 
\end{equation}
relating the Sommerfeld coefficient $\gamma$ to the static spin and charge susceptibilities $\chi_s$, $\chi_c$.

We characterise the LFL--HFL transition within this framework. We show that it is discontinuous at low temperature, forced by an algebraic constraint on the Hubbard contraction, and we argue on entropic grounds that this discontinuity terminates at a finite-temperature critical endpoint. This is the point of direct competition between the two metallic states, which we argue drives a breakdown of momentum-space coherence, giving rise to the $0+1$D bath. We present the resulting phase diagram in \figref{fig:phasediagram}. 

This $0+1$D description connects to, but is distinct from, the SYK \cite{SachdevYe1993, Kitaev2015} and DMFT \cite{Georges1996} frameworks. Unlike those, which respectively rely on infinite-range couplings or infinite spatial dimensions, we argue that a $0+1$D bath emerges in finite-dimensional lattices as a consequence of the LFL--HFL transition.

This dimensional reduction also aligns with the concept of local quantum criticality, proposed for the strange metal regime of cuprate and heavy-fermion systems as a temporal criticality accompanying the loss of a coherent momentum-space description \cite{Varma1989, Si2001}. While those frameworks were advanced to account for the anomalous transport of these systems, our approach offers an algebraic route to the dimensional reduction itself. In this work we focus on the thermodynamic properties of this $0+1$D description, leaving the question of transport and marginal phenomenology as an important open direction.

The paper is structured as follows. In \secref{sec:organising}, we review the canonical and Hubbard algebraic formulations of the electron, and discuss the transition between their respective metallic states. In \secref{sec:algebra}, we detail the exceptional Lie superalgebra $D(2,1;\alpha)$, and demonstrate how it contracts to yield the two distinct electronic formulations. In \secref{sec:eosm}, we present our central conjecture for the strange metal as an emergent $0+1$D superconformal bath, derive its exceptional thermodynamics, and map out the phase diagram of the LFL--HFL transition. In \secref{sec:micro}, we provide a microscopic framing, drawing connections to distinguished lattice models and the phenomenology of prominent material classes. Finally, we discuss broader implications and future directions in \secref{sec:discussion}, before concluding in \secref{sec:conclusion}.

\section{The electronic degree of freedom}
\label{sec:organising}

The core challenge of many-body quantum physics is the exponential complexity of interacting systems, and an efficient description relies upon identifying appropriate degrees of freedom. A well-chosen degree of freedom isolates the most physically relevant correlations, providing the foundation for an effective quasiparticle description.

Mathematically, a quantum degree of freedom is specified by the algebra it obeys. In Heisenberg's equations of motion, the dynamics of an operator $\mathcal{O}$ are governed by the commutator $[H, \mathcal{O}]$. In the presence of interactions, these equations cannot be closed exactly, and approximations are required. The choice of algebra guides the truncation of these equations, ultimately dictating the dispersive modes. Indeed, any perturbative approximation carries an inherent, often implicit, assumption: that the chosen algebraic framework encodes the system's dominant physics.

Here, we apply this principle to the electron \cite{Quinn2018, Quinn2021}. The Hilbert space of a single electron consists of four basis states:
\begin{equation}
\ket{\circ}=\ket{0}, \quad \ket{\downarrow}=c^\dagger_\downarrow\ket{0}, \quad \ket{\uparrow}=c^\dagger_\uparrow\ket{0}, \quad \ket{\bullet}= c^\dagger_\downarrow c^\dagger_\uparrow\ket{0}
\end{equation}
corresponding to an empty site, a singly occupied site with spin down or up, and a doubly occupied site, respectively. Depending on how transitions between these four states are decomposed, one arrives at two distinct algebraic formulations of the electronic degree of freedom, which characterise two distinct metallic states.

\subsection{Canonical: $\mathfrak{su}(1|1) \oplus \mathfrak{su}(1|1)$}
\label{sec:canonical}

The canonical formulation treats the electron as two independent fermion species expressed through the algebra,
\begin{equation}
\{c_\sigma,c^\dagger_{\sigma'}\}=\delta_{\sigma\sigma'},
\end{equation}
with $\sigma=\downarrow,\uparrow$ denoting the electron's spin. Formally, the relation $\{c,c^\dagger\}=1$ gives the Lie superalgebra $\mathfrak{su}(1|1)$, and so we denote this canonical formulation as $\mathfrak{su}(1|1) \oplus \mathfrak{su}(1|1)$. This can be viewed as grouping the four basis states as $\big\{ \ket{0};\ket{\downarrow}\big\}\otimes \big\{\ket{0};\ket{\uparrow}\big\}$.

The canonical algebra exhibits a $U(1)$ charge symmetry generated by the number operators, $n_\sigma = c^\dagger_\sigma c_\sigma$. Incorporating their commutation relations,
\begin{equation}
[n_\sigma,c^\dagger_{\sigma'}]=\delta_{\sigma\sigma'} c^\dagger_{\sigma}, \qquad [n_\sigma,c_{\sigma'}]=-\delta_{\sigma\sigma'} c_{\sigma}
\end{equation}
extends the underlying algebra to $\mathfrak{u}(1|1) \oplus \mathfrak{u}(1|1)$ (that is, the $n_\sigma$ are generators with non-zero supertrace, e.g. $\bra{0}n_\uparrow\ket{0}-\bra{\uparrow}n_\uparrow\ket{\uparrow}\neq 0$).

A defining characteristic of this canonical formulation is that the anti-commutator yields a scalar operator, the identity. Focusing on a single species, a non-interacting kinetic Hamiltonian, $H=\sum_{i,j} \varepsilon_{ij} c^\dagger_{i} c_{j}$, has linear action on the creation operators, $[H, c^\dagger_{i}] = \sum_j \varepsilon_{ij} c^\dagger_{j}$, which allows for the straightforward identification of single-particle modes with dispersion $\varepsilon_k$. Interactions can be incorporated through the standard techniques of many-body perturbation theory, which provides the microscopic foundation for describing conventional metals as Landau--Fermi liquids. In particular, the Dyson equation expresses the single-particle Green's function,
\begin{equation}\label{eq:G_can}
G_k = \frac{1}{\omega +\mu - \varepsilon_k - \Sigma_k},
\end{equation}
through the self-energy $\Sigma_k$, which encodes the scattering-induced dressing of the bare modes as quasiparticles.

A powerful organizing principle within this perturbative framework is the concept of conserving approximations, due to Kadanoff and Baym \cite{BaymKadanoff1961}. One defines the Luttinger--Ward functional, $\Phi[G]$, from which the self-energy is obtained as $\Sigma_k=\frac{\delta\Phi}{\delta G_k}$, and any approximation constructed in this way inherently respects all conservation laws. 

A further consequence is the Luttinger sum rule, which relates the volume enclosed by the Fermi surface to the electron density, independently of the strength of interactions. This follows directly from the existence of $\Phi[G]$ \cite{LuttingerWard1960}, and has been formalised non-perturbatively as a general topological constraint on a translationally-invariant metallic state \cite{Oshikawa2000, ElseThorngrenSenthil2021}.

\subsection{Hubbard: $\mathfrak{su}(2|2)$}
\label{sec:hubbardalg}

We now turn to the Hubbard formulation of the electron, which we cast through the Lie superalgebra $\mathfrak{su}(2|2)$ \cite{Beisert2008}. This casting exposes useful structures within the algebra, along with a continuous parameter $\kappa$ which encodes a central extension. We summarize the essential details here, deferring the full algebraic relations and the explicit mapping to Hubbard's operators to \appref{app:Hubbard}.

The Hubbard algebra mixes all four electronic basis states, and here it is natural to group these as follows: $\big\{ |\downarrow\rangle,|\uparrow\rangle;~|\circ\rangle,|\bullet\rangle \big\}.$ The algebra is generated by two $\mathfrak{su}(2)$ triplets, $\vec{s}$ acting on the spin doublet $\{ |\downarrow\rangle,|\uparrow\rangle\}$ and $\vec{\eta}$ acting on the charge doublet $\{ |\circ\rangle,|\bullet\rangle\}$, alongside eight fermionic generators $q_{\sigma\nu}$ and $q^\dagger_{\sigma\nu}$ (with $\sigma \in \{\downarrow,\uparrow\}$ and $\nu \in \{\circ,\bullet\}$) that act between the doublets. Importantly, the fermionic anti-commutation relations do not yield pure scalars, but instead close back onto the $\mathfrak{su}(2)$ generators:
\begin{equation}\label{eq:q_ac}
\begin{aligned} &\{ q_{\sigma\nu}, q^\dagger_{\sigma\nu}\} = \frac{1+\kappa^2}{4} - \kappa \left(\sigma s^z - \nu\eta^z\right), \\ &\{ q_{\downarrow\nu}, q^\dagger_{\uparrow\nu}\} = \kappa s^+, \qquad \{ q_{\sigma\circ}, q^\dagger_{\sigma\bullet}\} = -\kappa \eta^+, \\ &\{ q_{\uparrow\nu}, q^\dagger_{\downarrow\nu}\} = \kappa s^-, \qquad \{ q_{\sigma\bullet}, q^\dagger_{\sigma\circ}\} = -\kappa \eta^-, \\ &\{ q_{\sigma\nu}, q_{\sigma'\nu'}\} = \{ q^\dagger_{\sigma\nu}, q^\dagger_{\sigma'\nu'}\} = \frac{1-\kappa^2}{4}\epsilon_{\sigma\sigma'}\epsilon_{\nu\nu'}. \end{aligned}
\end{equation}
Here, when appearing as coefficients rather than labels, $\sigma$ and $\nu$ take values $+1$ for $\{\uparrow, \bullet\}$ and $-1$ for $\{\downarrow, \circ\}$, and $\epsilon$ is the antisymmetric tensor with convention $\epsilon_{\downarrow\uparrow}=\epsilon_{\circ\bullet}=1$. Together these generators form the Lie superalgebra $\mathfrak{su}(2|2)$, with an exceptional central extension for $\kappa \neq 1$. At $\kappa=0$, the non-scalar terms drop out of the anti-commutation relations, and the canonical relations are recovered. In \secref{sec:identkappa} we offer a physical interpretation of $\kappa$ in terms of correlated hopping for $0<\kappa\le1$. 

Just as the canonical algebra is extended by the number operator, this algebra admits a $U(1)$ extension via the generator $\theta$:
\begin{equation}\label{eq:theta}
\theta = \frac{1}{3}\big(\vec{\eta}\cdot\vec{\eta} - \vec{s}\cdot\vec{s}\big),
\end{equation}
which, acting on the basis states, weights the spin and charge doublets oppositely, $\theta|\sigma\rangle = -\tfrac14|\sigma\rangle$ and $\theta|\nu\rangle = \tfrac14|\nu\rangle$. Equivalently, $\theta = (n_\uparrow - \tfrac12)(n_\downarrow - \tfrac12)$, so $U\theta$ is the on-site Hubbard interaction, up to a shift of the chemical potential and constant. 
The operator commutes with both spin and charge, and its commutation relations with the fermions take a linear form:
\begin{equation}\label{eq:theta_q}
\begin{aligned} [\theta,q^\dagger_{\sigma\nu}] &= \frac{1+\kappa^2}{4\kappa} q^\dagger_{\sigma\nu} - \frac{1-\kappa^2}{4\kappa}\epsilon_{\sigma\sigma'}\epsilon_{\nu\nu'}q_{\sigma'\nu'} \\ [\theta,q_{\sigma\nu}] &= - \frac{1+\kappa^2}{4\kappa} q_{\sigma\nu} + \frac{1-\kappa^2}{4\kappa}\epsilon_{\sigma\sigma'}\epsilon_{\nu\nu'}q^\dagger_{\sigma'\nu'}. \end{aligned}
\end{equation}
Incorporating $\theta$ as a generator extends the algebra to $\mathfrak{u}(2|2)$, and it is this full algebra, evaluated at $\kappa=1$, that maps to Hubbard's graded projection operators.

The Hubbard and canonical algebras offer complementary bases for the operators acting on an electronic site. In terms of the canonical fermion, the $q^\dagger_{\sigma\nu}$ take a non-linear form,
\begin{equation}\label{eq:q_def}
\begin{aligned} &q^\dagger_{\sigma\circ} = \frac{1+\kappa}{2} c_{\bar{\sigma}} - \kappa n_\sigma c_{\bar{\sigma}}, \qquad q^\dagger_{\sigma\bullet} = \sigma \big( \frac{1-\kappa}{2} c^\dagger_\sigma + \kappa n_{\bar{\sigma}} c^\dagger_\sigma \big), \\ &q_{\sigma\circ} = \frac{1+\kappa}{2} c^\dagger_{\bar{\sigma}} - \kappa n_\sigma c^\dagger_{\bar{\sigma}}, \qquad q_{\sigma\bullet} = \sigma \big( \frac{1-\kappa}{2} c_\sigma + \kappa n_{\bar{\sigma}} c_\sigma \big), \end{aligned}
\end{equation}
with $\bar{\sigma}=-\sigma$. Remarkably, the inverse relations are linear:
\begin{equation}\label{eq:splitting}
c^\dagger_{\downarrow} = q_{\uparrow\circ}-q^\dagger_{\downarrow\bullet}, \qquad c^\dagger_{\uparrow} = q_{\downarrow\circ}+q^\dagger_{\uparrow\bullet},
\end{equation}
termed a splitting of the electron (as opposed to fractionalisation, which takes a product form \cite{SenthilSachdevVojta2003, Coleman2015}). This has a powerful consequence: the electronic Green's function decomposes into an exact linear combination of the Green's functions of the $q^\dagger_{\sigma\nu}$. Thus, a systematic computation of the $q^\dagger_{\sigma\nu}$ matrix Green's function offers an alternative route by which to characterise electronic correlations.

We organise this computation by distinguishing the scalar and non-scalar contributions to the fermionic anti-commutation relations, Eq.~\eqref{eq:q_ac}. From the scalar part we identify the bare modes, which are exact in $\kappa$. The non-scalar spin and charge terms induce correlations, which can be treated perturbatively. These terms encode scattering, but also influence the spectral weight of the Hubbard modes directly, tying it to the expectation values of spin and charge. It is instructive to contrast this with the canonical formulation, where the spectral weight of the canonical electron is fixed by $\{c_\sigma, c^\dagger_{\sigma'}\} = \delta_{\sigma\sigma'}$, and this identity is the precise origin of the 1 in the numerator of Eq.~\eqref{eq:G_can}. For the Hubbard formulation scattering and spectral weight cannot be cleanly disentangled, as these non-scalar terms contribute to both, reflecting an obstruction we return to in \secref{sec:hfl}.

The Hubbard interaction $U\theta$ plays a distinguished role here. Owing to the linear action of $\theta$ on the $q^\dagger_{\sigma\nu}$, $U$ enters the inverse $q^\dagger_{\sigma\nu}$ matrix Green's function as a frequency-independent term, on the same footing as $\mu$. We refer to this as an \emph{emergent chemical potential}. Unlike $\mu$, however, $U$ shifts the split components of $c^\dagger_\sigma$ oppositely. When translated to the electronic Green's function via the splitting relations, Eq.~\eqref{eq:splitting}, increasing $U$ thus results in a splitting of the electronic band in two (presented explicitly in \appref{app:bare_hubbard}). By contrast, within the canonical formulation $U$ enters through the self-energy $\Sigma_k$, dynamically dressing the bare modes.

A consequence of this band splitting is that these bare modes generally violate the electronic Luttinger sum rule. For example, when the bands are gapped, the Fermi level intersects just one band, and approaching half-filling from above, say, the Fermi surface shrinks to zero, whereas the electronic sum rule would dictate that it encloses half the Brillouin zone. We argue that this violation is intrinsic, as here the Hubbard modes, not the canonical electron, are the underlying degrees of freedom. We return to this point, and its relation to the general topological constraint, in \secref{sec:discussion}.

Consider now scattering of the Hubbard modes, over a uniform reference state. For small $U$, the split bare bands may overlap, and low-energy scattering is both intra-band and inter-band. As $U$ is increased, these bands pull apart, a gap opens in the electronic spectrum, and thereafter inter-band scattering is frozen out below the gap scale. In either regime, it follows from Landau's phase space argument that the decay rate of the gapless modes scales as $\omega^2$ at low energy, at leading order in the non-scalar terms. The resulting dispersive modes are thus long-lived quasiparticles at this level of approximation. Since the electronic Green's function is a linear combination of $q^\dagger_{\sigma\nu}$ matrix Green's functions, the corresponding poles carry over to the electron. Once the split bands are separated by a gap, perturbative scattering cannot merge them, and thus the violation of the electronic Luttinger sum rule persists.

The appearance of the spin and charge generators in the algebra shapes the fate of the Hubbard modes at low temperature. The $\vec{s}$ and $\vec{\eta}$ describe local moments, and in analogy with interacting spin systems, we generally expect these to order. This redistributes the modes' spectral weight directly, with a momentum dependence set by the ordering wavevector, without necessarily opening a gap. This expectation is consistent with the general topological constraint, which requires a uniform metallic ground state to satisfy a sum rule. The analogy with interacting spin systems is instructive: the disordered regime may be viewed as the counterpart of a paramagnetic state, with the distinction here being the presence of the fermionic generators in the algebra, giving rise to dispersive electronic modes.

\subsubsection{The Hubbard--Fermi liquid}\label{sec:hfl}

We argue that the Hubbard formulation of the electron characterises a legitimate state of electronic matter, a metal with long-lived quasiparticles organised by the $\mathfrak{su}(2|2)$ algebra. As this is fundamentally distinct from a Landau--Fermi liquid (LFL), we term this instead a Hubbard--Fermi liquid\footnote{There is a notational conflict here. The Heavy Fermi liquid, observed in local moment systems, shares the acronym HFL. A notable aspect of the $\mathfrak{su}(2|2)$ paradigm, however, is that it extends naturally to incorporate a local moment \cite{QuinnErten}, as detailed in \appref{app:kondo}. We take the perspective that if the HFL is a valid state of matter, then the Heavy Fermi liquid is an instance of it.} (HFL). 

Hubbard operators, and related approaches, have long been employed to study the interacting electron problem \cite{Hubbard1965, Roth1969, Zaitsev1976, ManciniAvella2004, OvchinnikovValkov2004}. There is an important distinction to make however. These works treated Hubbard operators as a computational basis for computing the electronic Green's function, with violation of the electronic Luttinger sum rule an artifact of an uncontrolled approximation. We argue that by regarding the Hubbard algebra as a true degree of freedom, the violation is an intrinsic feature that distinguishes the HFL from the LFL.

There is an obstacle to establishing the HFL, however. It is not straightforward to develop a systematic perturbative expansion which respects conservation laws order by order. Within the canonical formalism, the anti-commutator yields a scalar identity, which guarantees the applicability of Wick's theorem and enables the perturbative construction of the Luttinger--Ward functional $\Phi$. The existence of $\Phi$ rests upon $\frac{\delta\Sigma_k}{\delta G_{k'}}=\frac{\delta\Sigma_{k'}}{\delta G_k}$, which is a flatness condition for $\Phi$. In contrast, the Hubbard anti-commutation relations yield non-scalar terms, the spin and charge generators. While in the limit $\kappa\to0$ the algebra flattens, for finite $\kappa$ these non-canonical contributions render Wick's theorem inapplicable, and their inherent curvature obstructs the definition of a corresponding $\Phi$-functional. This raises the question of how to formulate a systematic conserving expansion for the non-canonical degree of freedom.

This obstruction to a clean perturbative expansion is not unique to the Hubbard electron. It is instructive again to draw a parallel with interacting spin systems, where the physical operators obey the $\mathfrak{su}(2)$ spin algebra. With respect to a polarised reference state $\ket{\uparrow}$, the algebra is cast as: 
\begin{equation}
[a,a^\dagger]=1-\tfrac{1}{S} n,\quad [n,a^\dagger]=a^\dagger,\quad [n,a]=-a,
\end{equation}
with $a = \tfrac{1}{\sqrt{2S}} S^+$, $a^\dagger = \tfrac{1}{\sqrt{2S}}S^-$, $n = S  - S^z$, and spin wave theory is formulated as an expansion about the large-$S$ limit. As above, no $\Phi$-derivable approximation scheme is known to exist. The standard technique, based on the Holstein-Primakoff mapping to canonical bosons \cite{HolsteinPrimakoff1940}, results in a mixing of symmetry sectors across different orders in the $1/S$ expansion. While workarounds exist, each has its trade-offs, and a single coherent framework remains an open challenge \cite{Auerbach1994, Takahashi1987, KriegKopietz2019}. Nevertheless, spin-wave theory offers a controlled and quantitatively predictive description of magnetically ordered matter, the absence of a $\Phi$-functional notwithstanding.

We therefore proceed on the assumption that the HFL is also a legitimate state, with the obstruction a problem of method rather than of existence. In doing so, we will place the HFL within a broader algebraic structure, which clarifies both its origin and its relationship to the LFL.

\subsection{Competing metallic regimes}
\label{sec:competing}

We have argued that the canonical and Hubbard formulations of the electron are not merely complementary mathematical bases, but that they characterise two distinct metallic states. The canonical formulation underpins the LFL, where the charge carrier density is fixed by the Luttinger sum rule. The Hubbard formulation underpins the HFL, where the Hubbard interaction serves as an emergent chemical potential that splits the electronic band in two, and the electronic Luttinger count is generically violated.

This mismatch of the Luttinger count is a signature of the distinct nature of the two regimes. Within the LFL, the Luttinger count is a topological property of the Fermi surface, protected so long as the quasiparticles remain coherent. As the HFL Fermi surface violates the electronic Luttinger count, any transition from a LFL to a HFL requires a Fermi surface rearrangement. The two metals thus cannot be adiabatically connected: any path joining them at $T=0$ must be non-analytic.

The LFL--HFL transition is not confined to the infrared. The splitting of the electronic band in two requires a global redistribution of spectral weight, across the energy scale of the emergent chemical potential $U$ \cite{Eskes1991}. This distinguishes it from conventional transitions, in which the spectral rearrangement is confined to the low-energy vicinity of the Fermi surface.

This large-scale spectral rearrangement presents a theoretical hurdle. It cannot be captured perturbatively, neither by truncating the equations of motion of the canonical many-body perturbation theory, nor through an as-yet-unestablished non-canonical expansion. Instead, an overarching algebraic framework is required, and this is what we propose in \secref{sec:algebra}.

We can frame the location of the LFL--HFL transition by considering the coherence scales of the two metallic regimes. Deep within the LFL and HFL, each metal is protected by its own coherence scale, $T^*_{\text{LFL}}$ and $T^*_{\text{HFL}}$ respectively, below which the corresponding quasiparticle description is robust. As the system is tuned towards the LFL--HFL boundary, however, these scales compete, and we identify the transition regime with their intersection, $T^*_{\text{LFL}} \sim T^*_{\text{HFL}}$. We will argue in \secref{sec:eosm} that this occurs at finite temperature.

At temperatures above this intersection, the coherent quasiparticle of either metal is lost. We will argue that this is a critical regime, born of LFL--HFL competition, which realises an emergent superconformal symmetry, governed by the algebraic structure we now present in \secref{sec:algebra}.

\section{The exceptional Lie superalgebra $D(2,1;\alpha)$}
\label{sec:algebra}

In this section we detail the exceptional Lie superalgebra $D(2,1;\alpha)$, and demonstrate that it encompasses and extends the canonical and Hubbard formulations of the electron. First introduced in Kac's classification \cite{Kac1977}, it is the only simple Lie superalgebra with continuously deformable structure constants. We employ here a non-compact real form, in which one bosonic sector is the conformal algebra $\mathfrak{sl}(2,\mathbb{R})$, which underlies the emergent scale invariance of \secref{sec:eosm}. We detail the correspondence to the compact form in \appref{app:exceptional}, and also highlight some useful limits.

The algebra comprises the conformal triplet $D$, $E_+$, $E_-$, together with two bosonic $\mathfrak{su}(2)$ triplets $J^a_b$, $J^\alpha_\beta$, and two fermionic quartets $Q^{a\alpha}$ and $Q^\dagger_{a\alpha}$, with indices $a$ and $\alpha$ both taking values in $\{1,2\}$. The three sets of bosonic generators mutually commute and obey:
\begin{equation}
\begin{aligned} &[D, E_+] = E_+, \quad [D, E_-] = -E_-, \quad [E_+, E_-] = 2D,\\ &[J^a_b, J^c_d] = \delta^c_b J^a_d - \delta^a_d J^c_b, \qquad [J^\alpha_\beta, J^\gamma_\delta] = \delta^\gamma_\beta J^\alpha_\delta - \delta^\alpha_\delta J^\gamma_\beta. \end{aligned}
\end{equation}
The fermionic generators transform under the $\mathfrak{sl}(2,\mathbb{R})$ conformal generators as:
\begin{equation}
\begin{aligned} [D, Q^\dagger_{a\alpha}] &= \frac{1}{2} Q^\dagger_{a\alpha}, \qquad\qquad\qquad\; [D, Q^{a\alpha}] = -\frac{1}{2} Q^{a\alpha},\\ [E_+, Q^{a\alpha}] &= \epsilon^{ab} \epsilon^{\alpha\beta} Q^\dagger_{b\beta}, \qquad\qquad\; [E_+, Q^\dagger_{a\alpha}] = 0,\\ [E_-, Q^\dagger_{a\alpha}] &= \epsilon_{ab} \epsilon_{\alpha\beta} Q^{b\beta}, \qquad\qquad [E_-, Q^{a\alpha}] = 0, \end{aligned}
\end{equation}
and in the fundamental and conjugate representations of the two $\mathfrak{su}(2)$ factors:
\begin{equation}
\begin{aligned} [J^a_b, Q^{c\gamma}] &= \delta^c_b Q^{a\gamma} - \frac{1}{2}\delta^a_b Q^{c\gamma}, \qquad [J^a_b, Q^\dagger_{c\gamma}] = -\delta^a_c Q^\dagger_{b\gamma} + \frac{1}{2}\delta^a_b Q^\dagger_{c\gamma}\\ [J^\alpha_\beta, Q^{c\gamma}] &= \delta^\gamma_\beta Q^{c\alpha} - \frac{1}{2}\delta^\alpha_\beta Q^{c\gamma}, \qquad [J^\alpha_\beta, Q^\dagger_{c\gamma}] = -\delta^\alpha_\gamma Q^\dagger_{c\beta} + \frac{1}{2}\delta^\alpha_\beta Q^\dagger_{c\gamma}. \end{aligned}
\end{equation}
The anticommutators of the fermionic generators close onto the bosonic generators, weighted by structure constants $c_1$, $c_2$, $c_3$:
\begin{equation}\label{eq:Dqq}
\begin{aligned} &\{Q^{a\alpha}, Q^\dagger_{b\beta}\} = c_1 \delta_b^a \delta^\alpha_\beta D - c_2 \delta^\alpha_\beta J^a_b - c_3 \delta_b^a J^\alpha_\beta,\\ &\{Q^\dagger_{a\alpha}, Q^\dagger_{b\beta}\} = -c_1 \epsilon_{ab} \epsilon_{\alpha\beta} E_+, \qquad \{Q^{a\alpha}, Q^{b\beta}\} = c_1 \epsilon^{ab} \epsilon^{\alpha\beta} E_-, \end{aligned}
\end{equation}
which are constrained to obey:
\begin{equation}\label{eq:SJC}
c_1 + c_2 + c_3 = 0,
\end{equation}
as the requirement of the super-Jacobi identity. Because an overall rescaling of the supercharges rescales the structure constants uniformly, the algebra depends on a single continuous parameter, $\alpha=c_3/c_2$.

Due to the non-compact $\mathfrak{sl}(2,\mathbb{R})$ sector, unitary representations of this algebra are generally infinite-dimensional. We next show, however, that it is possible to contract the conformal sector and arrive at a four-dimensional representation in two distinct ways.

\subsection{Maximal contraction}
\label{sec:maxcontraction}

We contract all generators simultaneously via In\"on\"u--Wigner scaling \cite{InonuWigner1953}:
\begin{equation}
\tilde{D} = \lambda D, \qquad \tilde{E}_\pm = \lambda E_\pm, \qquad \tilde{J}^a_b = \lambda J^a_b, \qquad \tilde{J}^\alpha_\beta = \lambda J^\alpha_\beta, \qquad \tilde{Q}^{a\alpha} = \lambda^{1/2} Q^{a\alpha}, \qquad \tilde{Q}^\dagger_{a\alpha} = \lambda^{1/2} Q^\dagger_{a\alpha}.
\end{equation}
Taking $\lambda \to 0$, the commutators scale as $[\tilde{J}, \tilde{J}] \propto \lambda$ and $[\tilde{J}, \tilde{Q}] \propto \lambda$, vanishing entirely. Thus, the non-Abelian bosonic generators get flattened into commuting central charges, while the fermionic anti-commutators survive exactly:
\begin{equation}
\begin{aligned} &\{\tilde{Q}^{a\alpha}, \tilde{Q}^\dagger_{b\beta}\} = c_1 \delta_b^a \delta^\alpha_\beta \tilde{D} - c_2 \delta^\alpha_\beta \tilde{J}^a_b - c_3 \delta_b^a \tilde{J}^\alpha_\beta,\\ &\{\tilde{Q}^\dagger_{a\alpha}, \tilde{Q}^\dagger_{b\beta}\} = -c_1 \epsilon_{ab} \epsilon_{\alpha\beta} \tilde{E}_+, \qquad \{\tilde{Q}^{a\alpha}, \tilde{Q}^{b\beta}\} = c_1 \epsilon^{ab} \epsilon^{\alpha\beta} \tilde{E}_-, \end{aligned}
\end{equation}
leaving the coefficients $c_i$ untouched. We can now diagonalise the bosonic sectors, and setting $\tilde{E}_\pm = 0$, the brackets $\{\tilde{Q}, \tilde{Q}\}$ and $\{\tilde{Q}^\dagger, \tilde{Q}^\dagger\}$ vanish. Absorbing the continuous eigenvalues of the surviving central charges into the coefficients results in:
\begin{equation}
\{\tilde{Q}^{a\alpha}, \tilde{Q}^\dagger_{b\beta}\} = \hat{c}_1 \delta_b^a \delta^\alpha_\beta - \hat{c}_2 \delta^\alpha_\beta (\sigma^z)^a_b - \hat{c}_3 \delta_b^a (\sigma^z)^\alpha_\beta,
\end{equation}
where the new coefficients are unconstrained (the super-Jacobi identity here follows trivially from $[\tilde{J}, \tilde{Q}] = 0$). The right-hand side of this anti-commutator defines a diagonal $4 \times 4$ matrix in the basis of the operators $(a,\alpha) \in \{(1,1), (1,2), (2,1), (2,2)\}$:
\begin{equation}
\{\tilde{Q}, \tilde{Q}^\dagger\} = \text{diag}(\hat{c}_1 - \hat{c}_2 - \hat{c}_3, \;\; \hat{c}_1 - \hat{c}_2 + \hat{c}_3, \;\; \hat{c}_1 + \hat{c}_2 - \hat{c}_3, \;\; \hat{c}_1 + \hat{c}_2 + \hat{c}_3).
\end{equation}
Generically, these eigenvalues are all non-zero, resulting in four independent fermion pairs, yielding a minimal representation of dimension $2^4=16$. This can be reduced to $4$ by constraining the coefficients to decouple two of these pairs. Imposing unitarity, this requires setting either $\hat{c}_1 = |\hat{c}_2|$ with $\hat{c}_3=0$, or $\hat{c}_1 = |\hat{c}_3|$ with $\hat{c}_2=0$. Fixing the overall normalization to $|\hat{c}_i| = 1/2$, we thus obtain four possibilities for the two surviving canonical pairs, parameterized by $(\hat{c}_1, \hat{c}_2, \hat{c}_3)$:
\begin{equation}
\begin{aligned} (1/2, \, 0, \, -1/2)&:\quad \{\tilde{Q}^{a1}, \tilde{Q}^\dagger_{b1}\} = \delta^a_b,\\[6pt] (1/2, \, 0, \, 1/2)&:\quad \{\tilde{Q}^{a2}, \tilde{Q}^\dagger_{b2}\} = \delta^a_b,\\[6pt] (1/2, \, -1/2, \, 0)&:\quad \{\tilde{Q}^{1\alpha}, \tilde{Q}^\dagger_{1\beta}\} = \delta^\alpha_\beta,\\[6pt] (1/2, \, 1/2, \, 0)&:\quad \{\tilde{Q}^{2\alpha}, \tilde{Q}^\dagger_{2\beta}\} = \delta^\alpha_\beta. \end{aligned}
\end{equation}
That is, the algebra here contracts to $\mathfrak{su}(1|1)\oplus\mathfrak{su}(1|1)$ in four different ways.

\subsection{Partial contraction}
\label{sec:partialcontraction}

Alternatively, we can take advantage of the flexibility of the coefficients $c_i$ to contract just a single bosonic sector. Specifically, we contract the conformal sector via Beisert scaling \cite{Beisert2008}:
\begin{equation}
c_1 = \epsilon,\quad c_2 = -1,\quad c_3= 1-\epsilon,
\end{equation}
with the $\mathfrak{sl}(2,\mathbb{R})$ generators also scaled by $\epsilon$:
\begin{equation}
\tilde{D} = \epsilon D, \qquad \tilde{E}_+ = \epsilon E_+, \qquad \tilde{E}_- = \epsilon E_-,
\end{equation}
so that $\tilde{D}$, $\tilde{E}_\pm$ become central in the $\epsilon\to0$ limit. Here the fermionic generators are not scaled, but instead the scaling of $c_1$ balances the limit:
\begin{equation}
\begin{aligned} &\{Q^{a\alpha}, Q^\dagger_{b\beta}\} = \delta_b^a \delta^\alpha_\beta \tilde{D} + \delta^\alpha_\beta J^a_b - \delta_b^a J^\alpha_\beta,\\ &\{Q^\dagger_{a\alpha}, Q^\dagger_{b\beta}\} = -\epsilon_{ab} \epsilon_{\alpha\beta} \tilde{E}_+, \qquad \{Q^{a\alpha}, Q^{b\beta}\} = \epsilon^{ab} \epsilon^{\alpha\beta} \tilde{E}_-. \end{aligned}
\end{equation}
These, along with the surviving relations of the $J$ generators, define the centrally extended $\mathfrak{su}(2|2)$ algebra, or $\mathfrak{su}(2|2) \ltimes \mathbb{R}^3$. As above, generically this has a minimal representation of dimension $16$, but it shortens to the fundamental $4$-dimensional representation when the central charges satisfy the Bogomol'nyi-Prasad-Sommerfield (BPS) condition \cite{Beisert2007}:
\begin{equation}\label{eq:shortening}
\tilde{D}^2 - |\tilde{E}|^2 = 1/4,
\end{equation}
where $|\tilde{E}|^2 \equiv \tilde{E}_+^\dagger \tilde{E}_+= -\tilde{E}_+ \tilde{E}_-$. In this way, we obtain the Hubbard formulation of the electron. It will prove important that here $\alpha=\frac{c_3}{c_2}=-1$, an inescapable consequence of the super-Jacobi constraint, Eq.~\eqref{eq:SJC}, upon scaling $c_1$ to zero.

\subsection{Electronic interpretation}
\label{sec:electronicinterp}

We thus see that $D(2,1;\alpha)$ admits two contractions of the conformal sector, which lead to the two formulations of the electronic degree of freedom presented in \secref{sec:organising}.

To make this explicit, let us first focus on the appearance of the Hubbard algebra through the partial contraction. The two $\mathfrak{su}(2)$ algebras are identified as spin and charge by mapping the indices as follows,
\begin{equation}\label{eq:electron_id}
a,b \in \{1,2\} \leftrightarrow \{\downarrow, \uparrow\}, \quad \alpha,\beta \in \{1,2\} \leftrightarrow \{\circ, \bullet\},
\end{equation}
and their generators as projections onto the Pauli matrices $\sigma^i$ as $\tfrac{1}{2}\text{tr}(J^T \sigma^i)$:
\begin{equation}
s^i = \tfrac{1}{2} J^a_b (\sigma^i)^a_b, \quad \eta^i = \tfrac{1}{2} J^\alpha_\beta (\sigma^i)^\alpha_\beta,
\end{equation}
with $i \in \{x, y, z\}$. The fermionic generators then follow as $q_{\sigma\nu} = \sqrt{\kappa} Q^{\sigma\nu}$ and $q^\dagger_{\sigma\nu} = \sqrt{\kappa} Q^\dagger_{\sigma\nu}$, with the central charges taking values $\tilde{D} = \frac{1+\kappa^2}{4\kappa}$ and $\tilde{E}_\pm = \mp\frac{1-\kappa^2}{4\kappa}$, which obey Eq.\eqref{eq:shortening}.

For the maximal contraction we found four distinct limits that yield pairs of canonical fermions. From the identification above, we see that these correspond to the decoupling of one of the $\mathfrak{su}(2)$ sectors. This admits a natural physical interpretation. From the perspective of the Hubbard formulation of the electron, the natural reference state is at half-filling ($\braket{\vec{\eta}}=0$) and zero magnetic polarisation ($\braket{\vec{s}}=0$), while for a canonical electron it is at the extremes: an empty or full band, or a fully magnetically polarised state. Let us say a system realises a Mott insulator, then we expect the HFL regime to govern the vicinity of half-filling, and we can regard the two charge differentiated pairs, $\{\tilde{Q}^{\sigma\circ}, \tilde{Q}^\dagger_{\sigma'\circ}\} = \delta^\sigma_{\sigma'}$ and $\{\tilde{Q}^{\sigma\bullet}, \tilde{Q}^\dagger_{\sigma'\bullet}\} = \delta^\sigma_{\sigma'}$, as the canonical degrees of freedom for LFL regimes which extend from the band edges. Alternatively, if the system exhibits a correlation induced spin gap (occurring at negative $U$), then the two spin differentiated pairs, $\{\tilde{Q}^{\uparrow\nu}, \tilde{Q}^\dagger_{\uparrow\nu'}\} = \delta^\nu_{\nu'}$ and $\{\tilde{Q}^{\downarrow\nu}, \tilde{Q}^\dagger_{\downarrow\nu'}\} = \delta^\nu_{\nu'}$, map to the LFL regimes extending from the extremes of magnetic polarisation.

The identification of Eq.~\eqref{eq:electron_id} lifts naturally to the full $D(2,1;\alpha)$ algebra. We emphasise, however, that away from the contracted limits, the fermionic generators, $Q^{\sigma\nu}$ and $Q^\dagger_{\sigma\nu}$, no longer admit a strict electronic interpretation, as their anti-commutation relations yield the generators of the non-compact $\mathfrak{sl}(2,\mathbb{R})$.

In the following section we will interpret this as an emergent scale invariance, manifesting a dissolution of the electron, and enabling a transition between the canonical and Hubbard formulations of the electronic degree of freedom.

\section{The strange metal and its exceptional thermodynamics}
\label{sec:eosm}

In \secref{sec:organising} we presented the canonical and Hubbard forms of the electron as two distinct ways to characterise metallic behaviour, the LFL and HFL.  In \secref{sec:algebra}, we showed that the exceptional Lie superalgebra $D(2,1;\alpha)$ admits contractions of its conformal symmetry in two distinct limits, which map precisely onto these two electronic algebras. We now argue that $D(2,1;\alpha)$ provides the organising principle for the LFL--HFL transition, with its conformal $\mathfrak{sl}(2,\mathbb{R})$ sector governing an emergent temporal dynamics, enabling a reorganisation of the electronic spectrum. This is our conjecture for the exceptional origin of the strange metal: it is a manifestation of an emergent superconformal symmetry which characterises the competition between the two metallic states.

The emergence of $D(2,1;\alpha)$ symmetry presents strict kinematic constraints. Its conformal sector, $\mathfrak{sl}(2,\mathbb{R})$, is the conformal algebra of a zero-dimensional space, with purely temporal scale invariance. Because the superalgebra does not contain spatial generators, it cannot organise coherent dispersion. Moreover, no single site can realise the symmetry. Non-trivial unitary representations of $\mathfrak{sl}(2,\mathbb{R})$ are necessarily infinite-dimensional, while a single electronic site has just four states. Consequently, the electron must merge with its local environment, giving rise to a description of the system as a $0+1$D bath with a continuous spectrum in the thermodynamic limit.

In this section we examine the thermodynamics of this $0+1$D superconformal system, and will show that the resulting framework offers a self-consistent description of the LFL--HFL transition. In particular, it enables us to view the dimensional reduction to $0+1$D from the perspective of the metallic states, which we discuss in \secref{sec:momentumbreakdown}.

\subsection{Thermodynamics of the $0+1$D bath}
\label{sec:bath}

The thermodynamics of a $0+1$D conformal bath is determined by its pattern of symmetry breaking, following the Schwarzian effective action methodology developed for the SYK family of models \cite{Kitaev2015, MaldacenaStanford2016, FuGaiottaMaldacena2017}. In general, this works as follows. In the strict infrared limit, the system is invariant under an infinite-dimensional super-reparameterization symmetry, which is spontaneously broken down to its global supergroup by the ground state. Moving away from the infrared limit, finite energy scales explicitly break the full super-reparameterization symmetry, lifting the corresponding zero-modes into pseudo-Goldstone bosons and their fermionic partners.

For the bath at hand, the super-reparameterization symmetry is the super-Virasoro extension of $D(2,1;\alpha)$ \cite{GoddardSchwimmer1988, SevrinTroostVanProeyen1988}, known in the high-energy literature as the large $\mathcal{N}=4$ superconformal algebra\footnote{
    In the high-energy literature, the adjective \emph{large} distinguishes it from the \emph{small} $\mathcal{N}=4$ superconformal algebra: the super-Virasoro extension of $\mathfrak{psu}(1,1|2)$, possessing a single $\mathfrak{su}(2)$ current. The small algebra appears in our discussion of Mott criticality in \secref{sec:mottcriticality}. 
}, which we present in \appref{app:vira}. Its bosonic sector is
\begin{equation}
\text{Diff}(S^1) \ltimes \big( L(SU(2)) \times L(SU(2)) \times L(U(1)) \big),
\end{equation}
with $\text{Diff}(S^1)$ the group of mappings from the circle to itself, and $L(G)$ the loop group of mappings from the circle to $G$. The two $L(SU(2))$ factors are the spin and charge current algebras, and $L(U(1))$ is a decoupled $U(1)$ current which serves to linearise the algebra\footnote{$D(2,1;\alpha)$ also admits a minimal super-Virasoro extension which is non-linear \cite{GoddardSchwimmer1988}. An attempt to construct the corresponding Schwarzian effective action yields trivial dynamics \cite{GalajinskyMasterov2021}. The non-linearity also obstructs the fermionic localisation of the path integral, employed for computing the one-loop exact quantum correction in \secref{sec:oneloop}. We do not consider this non-linear variant here.}.
The fermionic sector comprises the four $D(2,1;\alpha)$ supercurrents, together with four free Majorana fermions associated with the decoupled $U(1)$ current. 

The ground state breaks this symmetry to the global supergroup $D(2,1;\alpha) \times U(1)$. The resulting pseudo-Goldstone bosonic modes are fluctuations of the temporal, spin and charge sectors, along with the decoupled $U(1)$ phase. 

We highlight that this additional $U(1)$ is distinct from the $U(1)$ generated by the Hubbard $\theta$ of \secref{sec:hubbardalg}, which acts non-trivially on the fermionic generators. The zero mode $Y_0$ of the $U(1)$ current $Y_n$ is central (\appref{app:vira}), and no physical source couples to it. The associated phase zero mode enters the one-loop counting of \secref{sec:oneloop}, while its winding sum contributes only exponentially small corrections (\appref{app:one-loop}).

\subsubsection{The effective action}

The partition function of this theory admits an exact solution, computed by Heydeman, Shi and Turiaci~\cite{HeydemanShiTuriaci2025}.  We focus our analysis on the thermodynamic observables of the bath, the free energy and the static spin and charge response, and  we proceed in stages: first the semiclassical saddle-point, and then the quantum correction. We express the partition function as,
\begin{equation}
Z = e^{S_0} \int \mathcal{D}[f, U_s, U_c, \ldots] \, e^{-S_{eff}},
\end{equation}
where $f$ denotes time reparameterisation, and the matrices $U_s(\tau), U_c(\tau) \in SU(2)$ parameterise fluctuations within the spin and charge sectors. The prefactor $e^{S_0}$ encodes the ground-state degeneracy, and we expect the residual entropy $S_0$ to be extensive, reflecting the intrinsically local nature of the $0+1$D bath. The ellipsis denote the fermionic modes and the decoupled $U(1)$ phase, which we do not consider for the semiclassical analysis as they contribute only at one loop.

The Schwarzian effective action for these bosonic pseudo-Goldstone fluctuations is \cite{HeydemanShiTuriaci2025, KozyrevKrivonos2022}:
\begin{equation}
S_{eff} = - M \int_0^{\beta} d\tau \left[ \{f, \tau\} + \lambda_s \text{Tr}(U_s^{-1} \partial_\tau U_s)^2 + \lambda_c \text{Tr}(U_c^{-1} \partial_\tau U_c)^2 
\right],
\end{equation}
where $\{f, \tau\} = \frac{f'''}{f'} - \frac{3}{2}\left(\frac{f''}{f'}\right)^2$ is the Schwarzian derivative, which penalises deviations from the conformal frame, and the dynamics of the $SU(2)$ currents are captured by Maurer-Cartan kinetic terms. The overall coefficient $M$ is an emergent scale that sets the stiffness of the bath.

The relative couplings $\lambda_s$ and $\lambda_c$ are constrained by $D(2,1;\alpha)$ invariance. Let us sketch this briefly. The action is fixed by contracting the currents with the invariant bilinear form $\langle X, Y \rangle \equiv \text{Str}(X Y)$, which satisfies the invariance condition: $\langle [X, Y\}, Z \rangle = \langle X, [Y, Z\} \rangle$. The bosonic terms, say $\langle D, D \rangle$, can be related through the fermionic anti-commutation relations: $\langle \{ Q, Q^\dagger\}, D \rangle = \langle Q, [Q^\dagger, D] \rangle$ which gives $\langle D, D \rangle =- \langle Q, Q^\dagger \rangle$. Obtaining similar equations for the $\langle J, J \rangle$ one obtains the relations $\lambda_s = - c_1/c_2$ and $\lambda_c = - c_1/c_3$, with the $c_i$ obeying the super-Jacobi constraint, Eq.~\eqref{eq:SJC}. That is, the couplings are constrained to obey:
\begin{equation}
\frac{1}{\lambda_s}+\frac{1}{\lambda_c}=1,
\end{equation}
and we parameterise this as follows:
\begin{equation}\label{eq:lambdas}
\lambda_s = 1+\alpha,\quad \lambda_c = 1 + \frac{1}{\alpha}.
\end{equation}
This matches the form of Ref.~\cite{HeydemanShiTuriaci2025} (their Eq.~(3.1)), save for the decoupled $U(1)$ whose thermodynamic response we do not consider.

There is a subtlety regarding the thermodynamic stability of the bath that is worthwhile highlighting. The conditions $\lambda_s=1+\alpha\ge 0$ and $\lambda_c=\frac{1+\alpha}{\alpha}\ge 0$, yield $\alpha>0$, the stability condition for the bath, together with an isolated point, $\alpha=-1$, where $\lambda_s=\lambda_c=0$ and the bath is degenerate. We return to this isolated point in \secref{sec:hfldiscontinuity}, where it appears at the emergence of the HFL.

\subsubsection{Semiclassical saddle-point}
\label{sec:thermo}

We now evaluate at the saddle-point, which gives the leading contribution within the conformal window:
\begin{equation}\label{eq:conformal_window}
\frac{1}{M}\ll T \ll T_{UV},
\end{equation} 
with $T = 1/\beta$, and $T_{UV}$ the ultraviolet scale at which the bath description is no longer valid. 
For the time-reparameterization mode, $f(\tau)$, the saddle point is the conformal transformation mapping the thermal circle to the zero-temperature line: $f(\tau) = \frac{\beta}{\pi}\tan\left(\frac{\pi\tau}{\beta}\right)$. Evaluating the Schwarzian derivative, $\{f, \tau\} = 2\pi^2/\beta^2$, results in the classical effective action:
\begin{equation}
S_{eff} = - M \int_0^\beta d\tau \left( \frac{2\pi^2}{\beta^2} \right) = -\frac{2\pi^2 M}{\beta} = -2\pi^2 M T.
\end{equation}
The free energy is $F = -T\log Z = -TS_0 + TS_{eff}=-TS_0 -2\pi^2 M T^2$, yielding the entropy
\begin{equation}
S = -\frac{\partial F}{\partial T} =S_0 + 4\pi^2 M T,
\end{equation}
and specific heat,
\begin{equation}
C_v =- T \frac{\partial^2 F}{\partial T^2} = 4\pi^2 M T.
\end{equation}
From the linear-in-$T$ scaling of $C_v$, we identify the Sommerfeld coefficient as $\gamma = 4\pi^2 M$.

To determine the static spin and charge susceptibilities, we introduce external background source fields via minimal coupling $D_\tau = \partial_\tau - A_\tau$. Let us focus on spin, and take the source $A_\tau^s = h_s s^z$. Parameterising the fluctuation around the $z$-axis as $U_s(\tau) = e^{i\phi_s(\tau) s^z}$, the spin contribution to the action is:
\begin{equation}
S_{eff}^{s} =  \frac{\lambda_s M }{2} \int_0^{\beta} d\tau\, (\partial_\tau \phi_s + i h_s)^2.
\end{equation}
This is the action of a compact quantum rotor, with the phase obeying the winding boundary condition $\phi_s(\beta) - \phi_s(0) = 4\pi m$ on the thermal circle, $m\in\mathbb{Z}$. In the semiclassical regime the partition sum is dominated by the trivial sector ($m=0$), yielding the source-dependent $\Delta F = -\tfrac{1}{2}\lambda_s M h_s^2$. The temperature dependence cancels, giving a constant, Pauli-like static spin susceptibility:
\begin{equation}
\chi_s = -\frac{\partial^2\Delta F}{\partial h_s^2} = \lambda_sM =(1+\alpha) M.
\end{equation}
The static charge susceptibility follows identically for $A_\tau^c = h_c \eta^z$:
\begin{equation}
\chi_c = -\frac{\partial^2 \Delta F}{\partial h_c^2} = \lambda_cM =\frac{1+\alpha}{\alpha}M.
\end{equation}

The non-universal stiffness $M$ can be eliminated by introducing the Wilson ratios:
\begin{equation}
\frac{\chi_s}{\gamma} = \frac{1+\alpha}{4\pi^2},\quad \frac{\chi_c}{\gamma} = \frac{1+\alpha}{4\pi^2\alpha},
\end{equation}
which allow for direct experimental determination of $\alpha$. Moreover, combining these yields an exceptional thermodynamic relation:
\begin{equation}\label{eq:ETR}
\frac{1}{\chi_s} + \frac{1}{\chi_c} = \frac{4\pi^2}{\gamma},
\end{equation}
a direct consequence of the super-Jacobi constraint, Eq.~\eqref{eq:SJC}. This constitutes a parameter-free experimental prediction for the $D(2,1;\alpha)$ strange metal, at leading order in the intermediate temperature regime of Eq.~\eqref{eq:conformal_window}. Restoring the fundamental physical constants ($k_B, g, \mu_B, e$), and adjusting for $n=2\eta^z$, we obtain Eq.~\eqref{eq:ETR_intro} of the Introduction.

\subsubsection{One loop correction}
\label{sec:oneloop}

The semiclassical approximation receives corrections from the quantum fluctuations of the soft modes. As shown by Stanford and Witten \cite{StanfordWitten2017}, the Schwarzian path integral localises and is one-loop exact by the Duistermaat--Heckman formula (this requires the super-reparameterisation algebra to be linear, which is ensured here by the decoupled $U(1)$). In the general case, the temperature dependence of the partition function for a super-Schwarzian theory takes the form \cite{HeydemanShiTuriaci2026}:
\begin{equation}
    Z(\beta) \propto \left(\frac{M}{\beta}\right)^p e^{\frac{2\pi^2 M}{\beta}},
\end{equation}
with $p = \frac{N_b-N_f}{2}$, where $N_b$ and $N_f$ are the numbers of bosonic and fermionic generators of the global supergroup. This gives a logarithmic in $T$ correction to the entropy,
\begin{equation}\label{eq:entropy_oneloop}
S =S_0 + 4\pi^2 M T + p\log(MT) + p,
\end{equation}
and sets a constant offset to the specific heat,
\begin{equation} \label{eq:Cv_oneloop}  
C_v = 4\pi^2 MT + p.
\end{equation}
The logarithmic divergence of the entropy at low-$T$ is pre-empted by the edge of the conformal window, $T\sim1/M$, where the quantum correction becomes comparable to the saddle contribution and the expansion breaks down. For our case the global supergroup is $D(2,1;\alpha)\times U(1)$, with $N_b=9+1$ bosonic and $N_f=8$ fermionic zero modes\footnote{The four fermions of the decoupled $U(1)$ linearise the algebra but are not generators of the global supergroup, and so do not enter the count.}, yielding $p=1$. We will return to this counting of modes later in \secref{sec:mottcriticality}, where we discuss Mott criticality and obtain a different constant offset to the specific heat.

Turning to the static susceptibilities, we obtain the one-loop correction from the exact source-dependent partition function~\cite{HeydemanShiTuriaci2025}:
\begin{equation}\label{eq:chi_oneloop}
\chi_s = (1+\alpha)M + \frac{1}{T}\Big(\frac16-\frac{1}{\pi^2}\Big),
\qquad
\chi_c = \frac{1+\alpha}{\alpha}M + \frac{1}{T}\Big(\frac16-\frac{1}{\pi^2}\Big),
\end{equation}
as detailed in \appref{app:one-loop}.
The correction is Curie-like with a universal coefficient: identical for spin and charge, and independent of $\alpha$. Similarly to the entropy, the low-$T$ divergence is again pre-empted by the window edge, $T\sim1/M$. We also note that the mixed susceptibility vanishes to this order: $\chi_{sc} = \frac{1}{\beta} \frac{\partial^2 \log Z}{\partial  h_s \partial h_c}\big|_0 = 0$.

These one-loop contributions carry over to the exceptional thermodynamic relation, Eq.~\eqref{eq:ETR}, giving a $O(1/MT)$ correction, which is suppressed for $T\gg1/M$.

\subsubsection{Charge symmetry breaking}
\label{sec:doping}

So far we have focused on an idealised, manifestly $D(2,1;\alpha)$ symmetric, conformal bath. We have motivated this, however, by the LFL--HFL transition, which we generally expect to occur away from half-filling. Doping away from half-filling explicitly breaks the charge $SU(2)$, and here we trace the consequences.

We focus on the charge kinetic term, including a chemical potential $\mu$ measured from the particle-hole symmetric point:
\begin{equation}
S_{eff}^c = - \lambda_c M \int_0^\beta d\tau \,\text{Tr}\big(U_c^{-1} \partial_\tau U_c - \mu \sigma^z\big)^2,
\end{equation}
which explicitly breaks the charge symmetry, $SU(2)_c \to U(1)_c$, lifting the degeneracy of the charge multiplet. As all fermionic modes are charged under $U(1)_c$, they acquire a charge gap of order $\mu$.

For $T \gg \mu$, thermal fluctuations overwhelm this gap and the bath retains its scale invariance. As the system cools to $T \sim \mu$, however, the fermionic modes become gapped, and the full emergent superconformal symmetry gets broken. 

As $T$ drops below this scale, the residual macroscopic entropy of the bath must be quenched. If the bath is destroyed above the coherence temperature scales of the LFL or the HFL (\secref{sec:endpoint}), then the remaining route is through the opening of a gap, via the spontaneous breaking of a residual symmetry: $U(1)_c$ or $SU(2)_s$.

\subsection{The fate of the electron}
\label{sec:fate}

We now consider the fate of the electron within this $0+1$D bath. We characterise this dynamically through the conformal symmetry at $T=0$. When an electron is injected into the scale-invariant ground state, its long-time dynamics are governed by the leading primary operator in the operator product expansion (OPE) of $c$. We denote the corresponding state as $|\Delta_e\rangle$, and its scaling dimension, $\Delta_e$, governs the asymptotic decay of the electronic Green's function:
\begin{equation}\label{eq:G_scaling}
G(\tau) = -\langle \mathcal{T} c(\tau) c^\dagger(0) \rangle \propto \frac{\text{sgn}(\tau)}{|\tau|^{2\Delta_e}}.
\end{equation}

To examine the relevance of spatial hopping, we consider the standard kinetic lattice action:
\begin{equation}
S_{kin} = \int d\tau \, H_{kin},\quad H_{kin} = -t \sum_{\langle i,j \rangle} \left( c^\dagger_i c_j + \text{h.c.} \right).
\end{equation}
Under $0+1$D RG scaling, time has scaling dimension $[\tau] = -1$, and this dictates the scaling dimension of the hopping parameter:
\begin{equation}
[t] = 1 - 2\Delta_e,
\end{equation}
with corresponding dimensionless coupling $\hat{t}(T) = t\,T^{2\Delta_e-1}$.

We obtain a bound on $\Delta_e$ from the anti-commutator of the $D(2,1;\alpha)$ fermionic charges, imposed by unitarity:
\begin{equation}
\begin{aligned} 0 \leq \bra{\Delta_e}\{Q^{\sigma\nu}, Q^\dagger_{\sigma\nu}\}\ket{\Delta_e} &= c_1 \bra{\Delta_e} \Big( D + \frac{1}{\lambda_s} J^\sigma_{\sigma} + \frac{1}{\lambda_c} J^\nu_{\nu} \Big) \ket{\Delta_e}\\ &= c_1 \Big(\Delta_e + \frac{m_s}{\lambda_s} + \frac{m_c}{\lambda_c} \Big), \end{aligned}
\end{equation}
with $c_1>0$. For the physical electron, the spin and charge quantum numbers are $m_s = \pm \frac{1}{2}$ and $m_c = \pm \frac{1}{2}$, and the inequality must hold for each.
For $\lambda_s, \lambda_c > 0$, optimising for the tightest inequality, and employing the super-Jacobi constraint, Eq.~\eqref{eq:SJC}, yields the BPS bound:
\begin{equation}\label{eq:bps_bound}
\Delta_e \geq \frac{1}{2}.
\end{equation}

At the bound, $\Delta_e = \tfrac12$, the expectation value above vanishes, and the corresponding supercharge annihilates the state. The electron multiplet shortens, locking the value $\Delta_e = \tfrac12$. Here the Green's function scales as $G(\tau) \propto \tau^{-1}$, indicating a stable mode. Saturation of the bound thus marks the boundary of the bath: here hopping is marginal, $[t]=0$, permitting a dimensional crossover out of the $0+1$D description to a coherent LFL.

By contrast, for $\Delta_e > \tfrac12$ the electron multiplet remains long. The anomalous scaling of the Green's function, Eq.~\eqref{eq:G_scaling}, yields the local spectral function $A(\omega) \propto |\omega|^{2\Delta_e-1}$, which vanishes at low frequency, reflecting a dissolution of the electronic quasiparticle. Hopping is irrelevant, $[t]<0$, and the electron is confined within the bath. The scaling also fixes the leading conductivity of the bath: the current operator carries dimension $[t] + 2\Delta_e = 1$, and the Kubo formula at second order in hopping gives
\begin{equation}
\sigma \;\propto\; \frac{e^2}{\hbar}\,\hat{t}(T)^2
\qquad\Longrightarrow\qquad
\rho \;\propto\; T^{\,2-4\Delta_e},
\label{eq:bath_rho}
\end{equation}
a resistivity that decreases with increasing temperature for any
$\Delta_e > \tfrac12$, and thus insulating.\footnote{
    It is instructive to contrast this with the SYK model, whose conformal window sits above the coherence scale set by the on-site $q$-body interaction, with fermion dimension $\Delta = 1/q$. This yields linear-in-$T$ resistivity at $q=4$, i.e.\ $\Delta = 1/4$ \cite{SongJianBalents2017, GuQiStanford2017}, below the BPS bound $\Delta_e \ge \tfrac12$, Eq.~\eqref{eq:bps_bound}. Thus that mechanism for conductivity within the bath is not available here.
}

\subsubsection{The Hubbard--Fermi liquid transition}
\label{sec:hfldiscontinuity}

Following this discussion for the canonical electron operator $c$, we now turn to the Hubbard formulation. 

The fermionic generators $q$ are governed by the identical BPS bound $\Delta_q \geq \frac{1}{2}$, as they carry the same spin and charge quantum
numbers as $c$. Expressing a kinetic term through the $q$,
$H_{kin} = -t \sum_{\langle i,j \rangle}
\big( q^\dagger_i q_j + \text{h.c.} \big)$,
the hopping parameter again scales as $[t] = 1 - 2\Delta_q$. As above, hopping is thus irrelevant for $\Delta_q > \frac{1}{2}$, and the electron
remains confined within the bath.

The exit, however, is fundamentally obstructed. Here saturation of the bound corresponds to a shortening of the Hubbard multiplet, and this is precisely the contraction to the Hubbard electron of \secref{sec:partialcontraction}. Unlike the canonical case, where the contraction is available at any value of $\alpha$ (\secref{sec:maxcontraction}), the contraction in the Beisert limit requires $\alpha=\frac{c_3}{c_2}\to-1$. This is the isolated degenerate point $\lambda_s=\lambda_c=0$, separated from the thermodynamically stable regime $\alpha > 0$ of the bath. We thus conclude that the HFL cannot emerge smoothly from the bath. The transition must be discontinuous at $T=0$, and it is natural to identify this with the rearrangement of the Fermi surface discussed in \secref{sec:competing}.

\subsubsection{The finite-temperature endpoint}
\label{sec:endpoint}

We argue that this $T=0$ discontinuity extends to a finite-temperature endpoint on entropic grounds. At intermediate temperature, thermal fluctuations permit a crossover from the bath to the HFL, and we denote this coherence scale as $T^*_{\rm HFL}$, as in \secref{sec:competing}. Likewise, we denote the coherence scale of the LFL as $T^*_{\rm LFL}$, and we can ask where these two crossover lines intersect. In the absence of ordering, the intersection point cannot extend down to zero temperature, as this would expose the residual entropy of the superconformal bath. Instead the crossover lines must meet at finite temperature, $T_e$, and we interpret this, the termination point of the superconformal bath, as the critical endpoint of a line that extends down to the $T=0$ discontinuity. 

If ordering does intervene first, then it pre-empts the endpoint, and we discuss this alternative possibility in \secref{sec:phasediagram}.

\subsubsection{Breakdown of momentum-space coherence}
\label{sec:momentumbreakdown}

Up to this point, we have conjectured the existence of the $0+1$D bath. In \secref{sec:fate} we showed that the bath is stable: within it, $\Delta_e>1/2$, and electronic hopping is an irrelevant perturbation. We now discuss this dimensional reduction from the perspective of the two metallic states. 

That the LFL--HFL transition is discontinuous, with a finite-temperature endpoint, has an important consequence: both Fermi liquids remain robust near the transition, with their respective coherence temperatures non-zero. The quasiparticle weight remains finite, and the transition is non-perturbative from either metallic state. Thus, we cannot expect to trace the dimensional reduction within either framework.

The transition is not of a conventional symmetry-breaking kind, as the two metallic states do not differ by a local order parameter. Instead, they are distinguished by the topology of their Fermi surfaces, which cannot be continuously deformed into one another while the quasiparticles remain coherent.

Near the transition, both formulations of the electron are reflected in the spectrum, but the low-energy weight is determined by which of the two has the higher coherence scale, with its quasiparticle defining the Fermi surface (\figref{fig:phasediagram}(a)). Below $T_e$ the transition is discontinuous, and the Fermi surface rearranges across it. At $T_e$ these two regimes meet, but the two formulations cannot simultaneously govern the low-energy dynamics, as their Fermi surfaces are incompatible. We argue that, where their coherence mechanisms compete, neither dominates, and a coherent momentum-space description must break down. Here, only local structure remains to organise the spectrum, and our conjecture is that the local degrees of freedom arrange into representations of $D(2,1;\alpha)$, enabling the emergence of a scale-invariant $0+1$D bath above the junction of the two metallic regimes.

Whether the totality of the local electronic spectral weight enters the bath, or whether a component remains locked at the marginal value $\Delta_e = 1/2$ (\secref{sec:fate}), we leave as an open question.

\subsection{Phase diagram}
\label{sec:phasediagram}

\begin{figure}[t]
\centering
\includegraphics[width=0.85\linewidth]{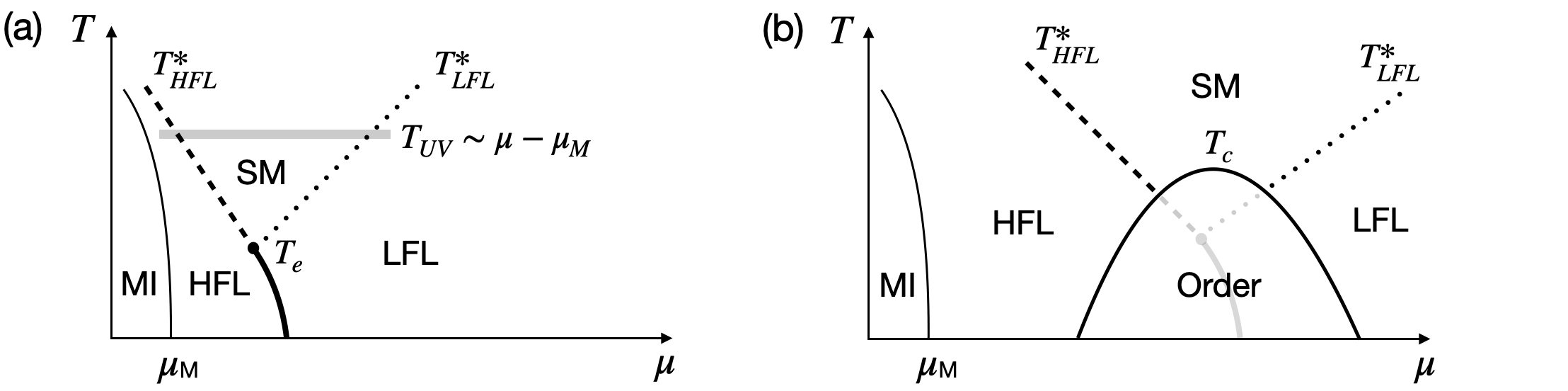}
\caption{Phase diagram of the LFL--HFL transition. The strange metal (SM) is the superconformal bath connecting the two Fermi liquids. The bold solid line is the discontinuous LFL--HFL transition up to the critical endpoint $T_e$, the dotted line is the generic SM-LFL crossover $T_{\rm LFL}^*$, the dashed line is the SM--HFL crossover of the HFL discontinuity $T_{\rm HFL}^*$, and the thin solid line is the boundary of the Mott insulator (MI), from which the HFL extends. The SM's residual entropy must be quenched by $T=0$, and two temperature scales compete to determine the outcome: $T_e$ and $T_c$, the latter being the temperature of the leading ordering instability of the SM (e.g.\ $T_c\sim \mu$, when finite doping breaks the charge $SU(2)$, \secref{sec:doping}). (a) $T_e>T_c$: the SM reaches the discontinuous LFL--HFL transition before an ordering instability appears, and the critical endpoint is exposed. (b) $T_c>T_e$: the bath orders first, and the endpoint is masked (displayed in gray). In addition, (a) also illustrates the approach to a subtle limit of our formalism, where the HFL phase shrinks away and the critical endpoint merges with the Mott boundary. The UV scale, $T_{UV}\sim \mu - \mu_M$ (bold gray line in (a)), drops to zero here, closing the conformal window of the strange metal from above, and giving rise to a distinct regime of Mott criticality (\secref{sec:mottcriticality}).}
\label{fig:phasediagram}
\end{figure}

We now map out the phase diagram of the LFL--HFL transition, which we present in \figref{fig:phasediagram}. We identify the superconformal regime connecting the two Fermi liquids as the strange metal (SM). This occupies an intermediate temperature range, $T_{IR}<T<T_{UV}$, with $T_{IR}>0$ as the residual entropy of the bath must be quenched before $T=0$, and the UV scale is that which resolves the spectral difference between LFL and HFL regimes. In the vicinity of the Mott boundary, this UV scale is set by $T_{UV}\sim \mu - \mu_M$, as in \figref{fig:phasediagram}(a).

The fate of the bath at low temperature is governed by the competition of two scales, as its residual entropy must be quenched by $T=0$. The first is $T_e$ itself, and the second is $T_c$, the temperature of the leading ordering instability of the SM bath. The outcome is determined by which is higher. When $T_e>T_c$, the SM reaches the endpoint before an ordering instability appears, leaving the endpoint exposed. Alternatively, when $T_c>T_e$, the bath orders at a temperature above the endpoint, masking the discontinuous transition. 

An interesting, yet subtle, limit is when the system is tuned so that the HFL phase shrinks away, resulting in the finite-temperature endpoint merging with the Mott insulator boundary. As the charge susceptibility diverges as the Mott gap is approached, we identify this limit as $\alpha\to0$. As a result, $\lambda_c\to\infty$, the charge decouples, and the thermodynamic relation reduces to $\frac{1}{\chi_s} = \frac{4\pi^2 }{\gamma}$. However, the temperature range over which this regime is valid shrinks to zero, as the UV scale, $T_{UV}\sim \mu - \mu_M$, drops down into the IR. Instead, a distinct critical regime emerges as the endpoint merges with the Mott transition, and we will return to discuss this Mott criticality in \secref{sec:mottcriticality}.

\section{Microscopic framing}
\label{sec:micro}

So far our discussion has been completely general, based on the assumption that the LFL and HFL exist as distinct metallic states, and our conjecture that the superconformal algebra $D(2,1;\alpha)$ governs the transition between them via an emergent superconformal bath.

We now turn to a microscopic perspective. We will focus in particular on the question of where the LFL--HFL transition may occur. As a function of band filling, we expect that the HFL regime, when it exists, is naturally anchored around half-filling, while the LFL regime extends from the band edges. We will consider whether there are limits in the parameter space of microscopic models where the LFL--HFL transition occurs right at half-filling, while a Mott gap is maintained, or alternatively where the HFL regime extends to the edges of its split bands. We will leverage exact results to suggest that the former occurs for the standard Hubbard model, while the latter occurs for the Hubbard-Shastry B (HSB) model.
We then take a material perspective, and frame this in the phenomenology of correlated insulators.

\subsection{Model Hamiltonian}
\label{sec:modelham}

We first consider a general single-orbital tight-binding model, and will then focus on a two-parameter subspace with some notable limits around which we will organise our discussion. Specifically we take a nearest-neighbour Hamiltonian on a bipartite lattice,
\begin{equation}\label{eq:H_gen}
H = \sum_{\langle i,j \rangle} T_{ij} + \sum_{\langle i,j \rangle} V^s_{ij} + \sum_{\langle i,j \rangle} V^\eta_{ij} + U \sum_j V^H_j.
\end{equation}
We denote the total number of lattice sites as $N$, and the coordination number of the lattice as $z$. We detail the terms as follows:
\begin{itemize}
\item Correlated hopping: the hopping term $T_{ij}$ goes beyond standard tight-binding by incorporating correlated hopping, where the amplitude depends on the occupancy of the sites involved. We decompose this as,
\begin{equation}
\begin{aligned}
T_{ij} &= T^\circ_{ij} + T^\bullet_{ij} + T^\pm_{ij} \\
T^\circ_{ij} &= -\sum_{\sigma=\downarrow,\uparrow} t(1-\lambda) \left( c^\dagger_{i\sigma} c_{j\sigma} + c^\dagger_{j\sigma} c_{i\sigma} \right) (1-n_{i\bar{\sigma}}) (1-n_{j\bar{\sigma}})\\
T^\bullet_{ij} &= -\sum_{\sigma=\downarrow,\uparrow} t(1+\lambda)\left( c^\dagger_{i\sigma} c_{j\sigma} + c^\dagger_{j\sigma} c_{i\sigma} \right) n_{i\bar{\sigma}} n_{j\bar{\sigma}}\\
T^\pm_{ij}&= -\sum_{\sigma=\downarrow,\uparrow} t_\pm \left( c^\dagger_{i\sigma} c_{j\sigma} + c^\dagger_{j\sigma} c_{i\sigma} \right) (n_{i\bar{\sigma}} - n_{j\bar{\sigma}})^2
\end{aligned}
\end{equation}
with $T^\circ_{ij}$, $T^\bullet_{ij}$, and $T^\pm_{ij}$ denoting hopping processes where the electrons of the opposite spin are absent, present at both sites, and present at only one site or the other, respectively. Conventional uncorrelated hopping is recovered for $t=t_\pm$, $\lambda=0$. The full correlated hopping term can alternatively be expressed through the $q^\dagger_{\sigma\nu}$ of Eq.~\eqref{eq:q_def}, where it takes a bilinear form:
\begin{equation}\label{eq:T_qq}
T_{ij} = -\sum_{\sigma=\downarrow,\uparrow}\sum_{\nu=\circ,\bullet} t_\nu \big( q_{i\sigma\nu}^{\dagger}q_{j\sigma\nu} + q_{j\sigma\nu}^{\dagger}q_{i\sigma\nu} \big),
\end{equation}
under the identification ($\nu$ takes values $-1,1$ for $\nu=\circ,\bullet$):
\begin{equation}\label{eq:CH_params}
t_\nu = \Big(\frac{2\nu}{1+\kappa^2}+\frac{\lambda}{\kappa}\Big)t,\quad \kappa=\sqrt{\frac{t-t_\pm}{t+t_\pm}}.
\end{equation}
\item Spin and charge interactions: cast with an anisotropic form,
\begin{equation}
V^s_{ij} = J_s^z s^z_i s^z_j + \frac{J_s^\perp}{2}(s^+_i s^-_j + s^-_i s^+_j),
\end{equation}
\begin{equation}
V^\eta_{ij} = J^z_\eta\eta^z_i \eta^z_j + \frac{J_\eta^\perp}{2}(\eta^+_i \eta^-_j + \eta^-_i \eta^+_j).
\end{equation}
\item Hubbard interaction: expressed through $\theta$ of Eq.~\eqref{eq:theta} as follows,
\begin{equation}
V^H_j = \theta_j = \big(n_{j\uparrow} - \tfrac{1}{2}\big)\big(n_{j\downarrow} - \tfrac{1}{2}\big),
\end{equation}
which is a particle-hole symmetric writing of the Hubbard interaction (the term linear in $n_\sigma$ amounts to a shift of the chemical potential).
\end{itemize}

\subsubsection{Identification of $\kappa$}
\label{sec:identkappa}

In presenting the Hubbard algebra in \secref{sec:hubbardalg} we did not specify a value for $\kappa$. For the microscopic model above, $\kappa$ encodes correlated hopping. This is parameterised through Eq.~\eqref{eq:CH_params}, which is fixed by the cancellation of off-diagonal hopping terms $q^\dagger q^\dagger$ and $qq$ in Eq.~\eqref{eq:T_qq}.

As this identification is set by the bare couplings, it would be interesting to take a renormalisation group (RG) perspective and ask how $\kappa$ behaves in the infrared (IR). Within the canonical formulation, correlated hopping is irrelevant in the IR for the LFL fixed point \cite{Shankar1994}. In contrast, the Hubbard formulation absorbs this interaction directly into the degree of freedom. To address renormalisation here, one could imagine formulating the path integral via coherent states of the Hubbard algebra, but as $\kappa$ enters this algebra directly, it is not immediately clear what an RG flow would mean in this case. An alternative is to express the Hamiltonian through the generators at fixed $\kappa$, and assess renormalisation there. In particular, $\kappa=1$ is a distinguished value, where the split bands obey their own sum rules (\secref{sec:discussion}). We leave this topic as a direction for future work.

\subsubsection{Distinguished models}
\label{sec:twoparam}

The general model possesses notable limits, which frame boundaries of the following two-parameter subspace parameterised by $(\kappa,u)$:
\begin{equation}
t=1, \quad \lambda=0, \quad J_s^z = J_s^\perp = -J_\eta^z = J_\eta^\perp = \frac{4\kappa}{1+\kappa^2}, \quad U = u + \frac{2z \kappa}{1+\kappa^2}.
\end{equation}
We consider the following one-parameter models\footnote{We restrict our attention to the upper-right quadrant in the $(\kappa,u)$ plane. We note, however, that the general Hamiltonian of Eq.~\eqref{eq:H_gen}, when particle-hole symmetric $\lambda=0$, is self-dual under the Shiba transformation \cite{Shiba1972},
\begin{equation}
c_{j\downarrow}^\dagger \leftrightarrow c_{j\downarrow}^\dagger, \quad c_{j\uparrow}^\dagger \leftrightarrow (-1)^j c_{j\uparrow},
\end{equation}
which interchanges spin and charge, and maps the model parameters as follows,
\begin{equation}
\{t,t_\pm,J_s^z,J_s^\perp,J_\eta^z,J_\eta^\perp,U\}\leftrightarrow\{t,t_\pm,J_\eta^z,-J_\eta^\perp,J_s^z,-J_s^\perp,-U\},
\end{equation}
or for the two-parameter subspace, $\{\kappa,u\}\leftrightarrow \{-\kappa,-u\}$ (and $\alpha\leftrightarrow\alpha^{-1}$ for the strange metal description, \secref{sec:eosm}). As a result, the lower-left quadrant of the $(\kappa,u)$ plane is obtained via this exact mapping, under which the Mott gap becomes a correlation-induced spin gap. We anticipate, on the other hand, that the behaviour in the upper-left and lower-right quadrants is governed by the closing of one gap and the opening of the other, but we do not pursue that further here.}:
\begin{itemize}
\item $(0, u)$ - The Hubbard model
\item $(\kappa,0)$ - The Hubbard-Shastry B (HSB) model \cite{FrolovQuinn2012}
\item $(1,u)$ - The Essler-Korepin-Schoutens (EKS) model \cite{EsslerKorepinSchoutens1992}
\item $(\kappa,\infty)$ - The Heisenberg spin model at half-filling, and the $t$-$J$ model away from half-filling.
\end{itemize}
The non-interacting limit resides at the origin $(0,0)$, and the Hubbard and HSB models offer two distinct routes to depart from there.

In one dimension these boundary models are highly distinguished: they represent limits where scattering becomes elastic and factorises, rendering them solvable by Bethe ansatz \cite{EsslerFrahm2005} (for the $t$-$J$ model away from half-filling, only at $J=2t$ \cite{BaresBlatter1990}). Away from the non-interacting point, the ground state is Mott insulating, and a unified study of their spectra reveals a high-energy excitation branch that switches between the charge sector of the Hubbard model and the spin sector of the HSB model \cite{QuinnHidden2021}. Physics in one dimension, however, is highly specialised, as the constrained geometry generically enforces low-frequency spin-charge separation \cite{Haldane1981, Giamarchi2003}. In this work our focus is on general dimensions greater than one, and we argue that these models continue to provide useful reference points as dimensionality is increased. 

The Hubbard model is tractable also in the limit of infinite dimensions \cite{MetznerVollhardt1989, Georges1996}. This is the realm of DMFT, where the lattice gets mapped to a single electronic site coupled to a self-consistent bath, enabling numerically exact computation of the local electronic spectral function. 

The HSB model possesses a dynamical symmetry on a bipartite lattice in any dimension, which leads to exact, protected modes. For this we define the global operators,
\begin{equation}
q^\dagger_{0\sigma\circ} = \frac{1}{\sqrt{N}} \sum_j q^\dagger_{j\sigma\circ},\quad q^\dagger_{\pi\sigma\bullet} = \frac{1}{\sqrt{N}}\sum_j (-1)^j q^\dagger_{j\sigma\bullet},
\end{equation}
corresponding to momentum $\vec{0}$ and $\vec{\pi}$, and thus the extremes of the electronic spectrum. These obey,
\begin{equation}\label{eq:HSB_qmodes}
[H_{HSB},q^\dagger_{0\sigma\circ}] = z\,q^\dagger_{0\sigma\circ}, \quad
[H_{HSB},q^\dagger_{\pi\sigma\bullet}] = z\,q^\dagger_{\pi\sigma\bullet},
\end{equation}
which confirms them as exact, infinite-lifetime modes.

The EKS model arises in the $\kappa\to 1$ limit of the HSB model, where the Hubbard interaction commutes with the Hamiltonian,
\begin{equation}\label{eq:H_v_theta}
[H_{EKS}, \sum_j\theta_j]=0,
\end{equation}
with the consequence that the parameter $u$ does not drive correlations, but instead serves as a true chemical potential for the split generators:
\begin{equation}\label{eq:EKS_theta_q}
[\theta,q^\dagger_{\sigma\nu}] = \frac{1}{2} q^\dagger_{\sigma\nu}, \quad [\theta,q_{\sigma\nu}] = - \frac{1}{2} q_{\sigma\nu}.
\end{equation}
The protected modes of the HSB model are here preserved for all $u$:
\begin{equation}
[H_{EKS},q^\dagger_{0\sigma\circ}] = \big(z+\frac{u}{2}\big)\, q^\dagger_{0\sigma\circ}, \quad
[H_{EKS},q^\dagger_{\pi\sigma\bullet}] = \big(z+\frac{u}{2}\big)\, q^\dagger_{\pi\sigma\bullet}.
\end{equation}

\subsubsection{Mapping the transition}
\label{sec:mapping}

The two formulations of the electron, canonical and Hubbard, offer distinct quasiparticle descriptions (\secref{sec:organising}). A priori both are available, and it is a challenging task to identify which governs the low-energy behaviour. We do not attempt this in general, but instead offer a perspective based upon the distinguished models of \secref{sec:twoparam}.

Specifically, we consider the location of the LFL--HFL transition as a function of band filling. We focus on the parent metallic state, overlooking ordering instabilities. We argue that the Hubbard model is a limit where the HFL phase is fully suppressed and the LFL extends to the Mott boundary at half-filling, when one exists, and that the HSB and EKS models are the opposite extreme, where the LFL phase is fully suppressed and the HFL extends out to the band edges.

For the Hubbard model this is borne out in the infinite-dimensional limit. Single-site DMFT shows that once a Mott gap opens at half-filling, it is separated from the doped metallic state by a first-order transition, and this remains the case as the strength of the Hubbard interaction is increased \cite{Georges1996, KotliarLangeRozenberg2000}. Employed in finite dimensions, it approximates the self-energy as momentum-independent, describing a LFL that extends up to the discontinuous Mott boundary. This is consistent with the Hubbard model's kinetic term taking the canonical bilinear form ($\kappa=\lambda=0$). Within our framework, this discontinuous transition at half-filling corresponds to the finite-temperature endpoint $T_e$ having merged with the Mott boundary, as discussed at the end of \secref{sec:phasediagram}. We will return to this from a material perspective with a discussion of Mott criticality in \secref{sec:mottcriticality}.

Taking $\kappa>0$ introduces correlated hopping, and spin and charge interactions. Generically, we expect a HFL regime to open up as these are introduced, with the consequence that the LFL--HFL transition moves away from half-filling.

For the HSB model, the evidence that the HFL regime extends across the whole (split) band is its protected fermionic modes, Eq.~\eqref{eq:HSB_qmodes}. As these carry momenta $\vec{0}$ and $\vec{\pi}$, we identify them as modes at the edges of the band. Moreover, in the limit to the EKS model at $\kappa\to1$, and generally when $t_\pm=0$, it follows from Eqs.~\eqref{eq:H_v_theta} and \eqref{eq:EKS_theta_q} that the electronic spectrum is strictly split into two, with the Hubbard interaction shifting the split bands oppositely in the spectrum. We thus suggest that the HFL phase extends across the whole band for the HSB model. This has a notable consequence: the band remains split with a Mott gap for all $0<\kappa\leq 1$, with the gap only vanishing in the non-interacting limit, $\kappa\to0$. We distinguish this from the Hubbard model, where above one dimension the Mott gap only opens at finite $U$. We understand this as follows. For the Hubbard model the split bands are incoherent, overlapping for small $U$, and the Mott gap opens once these shift apart. By contrast, for the HSB model the splitting occurs directly at the centre of the electronic band, opening a gap for $\kappa>0$.  

This analysis also offers an interpretation of the HSB model's dynamical fermionic symmetry as another special limit of our framework. The edges of the band represent the dilute limits of electronic excitations. As the LFL--HFL transition approaches these extremes, the residual entropy of the $0+1$D bath ceases to be extensive, and the critical endpoint $T_e$ drops to zero. From the perspective of $D(2,1;\alpha)$, this corresponds to the limit $\alpha\to1$, where the algebra becomes $\mathfrak{osp}(4|2)$ (\appref{app:exceptional}). Based on this, we conjecture that the HSB model exhibits a zero-density quantum critical point \cite{Sachdev2011}.

\subsection{Material perspective}
\label{sec:material}

This framing may help to understand which microscopic interactions drive the stability of the HFL at finite doping. In turn, this influences the temperature of the ordering instability of the strange metal, as discussed in \secref{sec:doping}.

To ground this in a material context, it is instructive to view our framing through the lens of the Zaanen-Sawatzky-Allen (ZSA) classification of transition-metal compounds \cite{ZaanenSawatzkyAllen1985}. The ZSA scheme distinguishes correlated insulators based on the competition between two energy scales: the on-site Coulomb repulsion $U$ of the transition-metal $d$-orbitals, and the charge-transfer energy $\Delta$ required to move an electron from the surrounding ligand $p$-orbitals to the metal. Their competition leads to the following dichotomy:
\begin{itemize}
\item Mott-Hubbard insulator: $U < \Delta$. The low-energy excitations are confined to the transition metal sites, and Hubbard $U$ is the dominant interaction.
\item Charge-transfer insulator: $\Delta<U$. The low-energy excitations are hybridized metal-ligand states, resulting in an effective single-orbital description with correlated hopping amplitudes, along with local spin and charge interactions.
\end{itemize}
The Hubbard model is the paradigmatic model of a Mott-Hubbard insulator. Although the specific couplings of the HSB model are fine-tuned and do not map to measured material parameters, we offer the perspective that it can be viewed as a model of a theoretically idealised charge-transfer insulator.

\subsubsection{Mott criticality}\label{sec:mottcriticality}

We focus now on Mott-Hubbard insulators. These are expected to be well-described by the Hubbard model, and so within our framework the critical endpoint is merged with the Mott boundary. The critical behaviour here is termed Mott criticality, and the prevailing paradigm is a DMFT prediction of a mean-field, liquid-gas Ising universality class \cite{KotliarLangeRozenberg2000}.

The status of this prediction is, however, unsettled. In a canonical Mott-Hubbard insulator, Cr-doped V$_2$O$_3$, mean-field scaling is reported \cite{Limelette2003}, with the caveat that the critical endpoint is obscured by magnetic and structural ordering. Alternatively, in frustrated organic Mott insulators where this ordering is suppressed and the endpoint exposed, non-mean-field criticality is instead reported \cite{KagawaMiyagawaKanoda2005, FurukawaKanoda2015}, and the topic remains under debate \cite{TanDobrosavljevicRademaker2022}.

Our framework offers a fresh perspective. As discussed at the end of \secref{sec:phasediagram}, the critical endpoint being merged with the Mott boundary represents a subtle limit of our formalism, $\alpha\to0$, where the UV scale of the superconformal bath of \secref{sec:bath} drops down into the IR. 

We can also track this limit algebraically. The limit $\alpha\to0$ is $\lambda_c\to\infty$ where the charge sector becomes infinitely stiff, and at the level of $D(2,1;\alpha)$, the charge generators drop out of Eq.~\eqref{eq:Dqq}, and decouple from the algebra, reducing it to $\mathfrak{psu}(1,1|2)$ (\appref{app:exceptional}). We thus conjecture that Mott criticality has a $PSU(1,1|2)$ superconformal symmetry. Here the super-Virasoro extension is the small $\mathcal{N}=4$ superconformal algebra \cite{HeydemanIliesiu2021}. As for the strange metal, this critical regime is $0+1$D, and we interpret the dimensional reduction that occurs here as the Mott localisation of the electron.

At the level of the semi-classical saddle, this gives the same $\frac{1}{\chi_s} = \frac{4\pi^2 }{\gamma}$ relation as the $\alpha\to0$ limit of our strange metal analysis, \secref{sec:thermo}. The two critical regimes are distinguished, however, by their exact one-loop correction. Here, the global supergroup $PSU(1,1|2)$ has $N_b=6$ and $N_f=8$, yielding $p=-1$ for Mott criticality, in contrast to the $p=+1$ result for strange metal criticality in \secref{sec:oneloop}. Physically, this reflects that the quantum fluctuations of the charge and $U(1)$ modes have dropped out, shifting the offset to the specific heat downwards, $C_v = 4\pi^2 MT + p$.

Let us offer an interpretation of the disagreement with the DMFT result. In single-site DMFT, the local problem is partitioned: an electron, with its four-state basis, is coupled to a self-consistent $0+1$D mean-field bath, which represents the spatial environment, and the thermal fluctuations of this external bath render the critical endpoint mean-field. In contrast, within our framework there is no partitioning: the $0+1$D bath emerges from the local sites themselves, with the basis states the representation of the symmetry algebra, which are inherently infinite in number due to the non-compact conformal sector. Here there is no external environment to wash out the quantum coherence of the critical regime.

\subsubsection{Cuprates}\label{sec:cuprates}

The cuprates are the most-studied charge-transfer insulators, and are natural candidates for an experimental realisation of our general framework. We map here the correspondence with their phenomenology at a qualitative level.

The hole-doped cuprate phase diagram (e.g.~Fig.~1 of Ref.\ \cite{Keimer2015}) closely resembles \figref{fig:phasediagram}(b). The mapping is direct: the HFL is the parent metallic state of the pseudogap, with its coherence scale $T_{HFL}^*$ the pseudogap temperature $T^*$, the LFL is the conventional metal recovered at high doping, the superconformal bath connecting them is the strange metal, and the Mott boundary from which the HFL extends is the antiferromagnetic insulator.

In contrast to scenarios which conjecture a zero temperature quantum critical point \cite{Broun2008, ProustTaillefer2019}, our framework posits a discontinuous transition with a finite-temperature critical endpoint. As discussed in \secref{sec:doping}, finite doping drives an ordering instability, at a scale set by the chemical potential, $T_c\sim\mu$. As the transition lies close to the Mott boundary ($\alpha\to0$, \secref{sec:mottcriticality}), we expect $\alpha<1$, and so $\tfrac{\chi_c}{\chi_s}=\tfrac{1}{\alpha}>1$ in our conventions. We thus expect spontaneous symmetry breaking of the residual charge $U(1)_c$ as the driving instability, consistent with a superconducting dome masking the finite-temperature endpoint.

Turning to the electron-doped side of the phase diagram \cite{ArmitageGreene2010, Greene2020}, the same mapping holds: we again identify the pseudogap regime with the HFL and the strange metal with the superconformal bath. Here the Mott phase extends closer to optimal doping, and the pseudogap regime is compressed. Unlike the hole-doped side where the strange metal extends to the highest measured temperatures, on the electron-doped side it crosses over to $T^2$ resistivity as temperature is raised. Within our framework, this deviation from linear-in-$T$ resistivity is consistent with the closer proximity to the Mott boundary: the ultraviolet cutoff $T_{UV} \sim \mu - \mu_M$, illustrated in \figref{fig:phasediagram}(a), is lower on the electron-doped side.

An important feature, common to both sides, is evidence of Fermi-surface rearrangement between the two metallic states \cite{DoironLeyraud2007, Fang2022, He2019}. We interpret this as the LFL--HFL transition underlying the strange metal. 

The marginal-Fermi-liquid phenomenology of the strange metal itself lies beyond the present analysis, and we return to this in \secref{sec:discussion}.

Finally, a characteristic feature of the pseudogap regime is its propensity to order, without destroying the metal. As discussed in \secref{sec:hubbardalg}, this is also an inherent feature of the HFL. There is a point worth highlighting here related to whether the ordered HFL state obeys a sum rule (\secref{sec:discussion}). The topological arguments imply that it does~\cite{Oshikawa2000, ElseThorngrenSenthil2021}, upon enlarging the unit cell, and this requires an emergent conserved density, related to the pattern of order. We would thus expect such an HFL state to be fragile upon moving away from the corresponding commensurate filling, and this may have a bearing on the variety of orders observed within the pseudogap regime \cite{FradkinKivelsonTranquada2015}.

\section{Discussion}
\label{sec:discussion}

We have developed an electronic interpretation of the exceptional Lie superalgebra $D(2,1;\alpha)$, whose structure maps directly onto the strongly correlated electron problem. In this framework the strange metal is an effective $0{+}1$D superconformal bath governed by $D(2,1;\alpha)$, the canonical and Hubbard electronic degrees of freedom are the two contractions of its conformal sector, and the LFL and HFL are the metallic states that form upon the emergence of coherent dispersion. We also examined the limit where the HFL regime is suppressed, and traced the origin of Mott criticality.

Central to this framework is the LFL--HFL transition. It is of an unconventional kind: it is discontinuous, with a finite-temperature critical endpoint, yet there is no associated symmetry-breaking. The two metallic states are distinguished not by a local order parameter, but by the topology of their Fermi surfaces, each protected by a robust quasiparticle. Within our framework the discontinuity has an algebraic origin: the superconformal bath is thermodynamically stable for $\alpha>0$, whereas the Hubbard contraction is locked to the degenerate point $\alpha=-1$ by the super-Jacobi constraint, so the HFL cannot emerge continuously from the bath (\secref{sec:hfldiscontinuity}). The transition is thus exceptional in origin: it is driven by the continuous deformability of the structure constants, a unique property of $D(2,1;\alpha)$ among simple Lie superalgebras.

We focused our analysis on the thermodynamics of the superconformal bath, the critical regime above the LFL--HFL transition. In particular, we obtained a parameter-free experimental prediction, Eq.~\eqref{eq:ETR}, relating the Sommerfeld coefficient to the static spin and charge susceptibilities, which provides a falsifiable test of our framework. Potential testing grounds are the strange metal regimes of the cuprates \cite{Keimer2015} and heavy-fermion compounds \cite{Stewart2001}. While the thermal and spin properties of these materials have been extensively characterised, the precise measurement of the charge response has proven more challenging \cite{AbbamonteFink2025}. More recently, 2D moir\'e materials, such as magic-angle twisted bilayer graphene, have been shown to exhibit strange metal behaviour \cite{Cao2020}. Here the experimental difficulty is inverted: $\chi_c$ is accessible via local compressibility measurements \cite{ParkCao2021}, but $\gamma$ and $\chi_s$ are more challenging owing to the small sample volume.

Beyond direct experimental measurement, our framework can be investigated numerically. Prominent here are the DMFT family of methods \cite{Georges1996, Maier2005}. Exact in the limit of infinite dimensions, the approach offers a window onto the metallic, insulating and critical regimes within a single framework. Cluster extensions, which incorporate short-range spatial correlations, have been shown to describe a discontinuous finite-temperature transition away from half-filling \cite{SordiHauleTremblay2010}. While the thermodynamic variables are computable within this approach, we are not aware of a single study that simultaneously extracts all three ($\gamma$, $\chi_s$, $\chi_c$) for the critical regime. There is a caveat worth highlighting, however. As discussed in \secref{sec:mottcriticality}, single-site DMFT renders the critical endpoint mean-field. It would be interesting to determine whether cluster extensions can resolve the intrinsically local superconformal criticality predicted by our framework.

Our identification of the superconformal bath as the strange metal rests on its placement as the critical regime between the two metallic states, mirroring the cuprates (\secref{sec:cuprates}) and heavy-fermion compounds \cite{Stewart2001}. However, the characteristic features of the strange metal, the linear-in-temperature resistivity and the logarithmic corrections to the specific heat of marginal Fermi liquid phenomenology \cite{Varma1989}, lie beyond our analysis here. We have seen that the bath itself supports no current (\secref{sec:fate}), and the leading thermodynamics is exact (\secref{sec:oneloop}), with no such logarithm. Instead, we anticipate that these signatures originate at the marginal boundary of the bath, $\Delta_e = 1/2$, where electronic weight is not wholly dissolved (\secref{sec:momentumbreakdown}). Computing the resulting response is a defect problem \cite{AffleckLudwig1991}, in which each lattice site defines a defect condition on the bath, as in the framework of local quantum criticality proposed for these materials \cite{Si2001}, and we leave this as a direction for future work.

The Hubbard--Fermi liquid is a pillar of our framework. While we have argued that it is a legitimate state of electronic matter, there is not yet a settled framework for characterising it. We highlighted in \secref{sec:hfl} that the requirement of a conserving, $\Phi$-derivable perturbative scheme may be too stringent, due to the intrinsic curvature of the Hubbard algebra. As spin-wave theory faces the same obstruction, we view this as a problem of method rather than of existence. The development of a systematic framework for such quantum degrees of freedom thus remains an important outstanding direction.

We have emphasised violation of the electronic Luttinger sum rule as an intrinsic feature of the HFL. Let us elaborate on this in view of the general microscopic model of \secref{sec:modelham}. The bare Hubbard modes can be computed exactly, and the resulting electronic Green's function is provided in \appref{app:bare_hubbard}. It is instructive to first set $t_\pm=0$ (and, for simplicity, $\lambda=0$), so that $\kappa=1$ and $\sum_j\theta_j$ is conserved. The bare electronic Green's function is then:
\begin{equation}\label{eq:G_kappa_1}
G_{k\sigma}(\omega) =\frac{1}{2}\sum_{\nu=\pm} 
\frac{1}{\omega+\mu+\frac{\nu U}{2}-\frac{\varepsilon_k}{2}},
\end{equation}
with two poles of equal weight, split by $U$. These split bands are filled by $q^\dagger_{\sigma\circ}$ and $q^\dagger_{\sigma\bullet}$, rather than the canonical $c^\dagger_\sigma$, so that each band accommodates half the electronic weight, and their density fills twice the Fermi volume. The electronic Luttinger sum rule is thus violated, except at $U=0$ where the split bands coincide. Instead, each band carries its own conserved charge $\eta^z\pm\theta$, and obeys a corresponding sum rule\footnote{
    In the language of Ref.~\cite{ElseThorngrenSenthil2021}, there is an emergent $U(1)$ loop group symmetry when the Fermi level intersects a band. For a Landau--Fermi liquid this symmetry is generated by the electron density, and the sum rule counts electrons. Here the densities are of $\eta^z\pm\theta$, which distinguish the split bands.
}. 
Turning to general $0<\kappa<1$, the $(1-\kappa^2)$ terms of Eqs.~\eqref{eq:q_ac} and \eqref{eq:theta_q} become non-zero, and $\sum_j\theta_j$ ceases to be conserved. Here the bare Hubbard modes do not appear to satisfy any sum rule, and whether one emerges in the infrared, for example as a flow of $\kappa\to1$, is an open question. Alternatively, we may expect the system to order at low temperature, as anticipated on physical grounds in \secref{sec:hubbardalg}, for consistency with the topological arguments of Refs.~\cite{Oshikawa2000, ElseThorngrenSenthil2021}, which forbid a uniform metallic ground state that does not obey a sum rule.

The HFL differs from prominent frameworks for a metal violating the Luttinger sum rule. The fractionalised Fermi liquid, FL$^*$, factorises the electron, $c \sim f\,b$, accompanied by an emergent gauge field, and is typically formulated at the level of an effective field theory rather than a microscopic operator identity \cite{SenthilSachdevVojta2003}. In contrast, the splitting of the HFL is linear, e.g.~$c^\dagger_{\downarrow} = q_{\uparrow\circ}-q^\dagger_{\downarrow\bullet}$, and its excitations retain the electronic quantum numbers. Another prominent approach is the notion of Mottness, which ties the violation to zeros of the canonical Green's function \cite{Phillips2006}. Within our framework, the canonical quasiparticle remains robust right up to the LFL--HFL transition, beyond which the low-energy spectrum is organised by the Hubbard formulation, and zeros play no role.

Our framework extends naturally to Kondo lattice systems \cite{QuinnErten}, relevant for heavy-fermion compounds \cite{Stewart2001}. As detailed in \appref{app:kondo}, the local moment is incorporated by enlarging the $\mathfrak{su}(2|2)$ representation, with the conduction-electron $\vec{s}$ and local moment $\vec{S}$ entering through the combination $\vec{\Sigma} = \vec{s} + \vec{S}$. The Kondo coupling then contributes to the emergent chemical potential that splits the band, and the splitting of the electron carries over. In this setting we regard the Heavy Fermi liquid as an instance of the HFL, with the local moment entwined with the electron rather than free to order independently. We can thus view the heavy-fermion phase diagram from the perspective of the LFL--HFL transition, with the strange metal emerging from the competing coherence mechanisms at the Fermi-surface rearrangement.

Finally, while our proposal for the exceptional origin of the strange metal stands independent of holographic duality, we highlight that our framework has deep roots in the literature of this subject. The algebraic structures we employed have been developed within the study of integrability in the AdS/CFT correspondence \cite{BeisertOverview2012, Bombardelli2016}, notably Beisert's discovery that the exceptionally extended $\mathfrak{su}(2|2)$ algebra governs the integrability of the 1D Hubbard model, along with its relationship to $D(2,1;\alpha)$ \cite{Beisert2008}. In addition, the superconformal theories we invoke for the strange metal and Mott criticality have been studied in the context of near-extremal black holes \cite{HeydemanIliesiu2021, HeydemanShiTuriaci2025, HeydemanShiTuriaci2026}, and these results underlie our formulation and analysis of the $D(2,1;\alpha)$ effective action.

\section{Conclusion}
\label{sec:conclusion}

We have developed a unified algebraic framework for the strongly correlated electron problem, encompassing the LFL and HFL metallic states, the strange metal, and Mott criticality. The organising structure is the exceptional Lie superalgebra $D(2,1;\alpha)$: the canonical and Hubbard electronic degrees of freedom are its two contractions, governing the LFL and HFL respectively, and the strange metal is the regime between the two, where the uncontracted symmetry is realised as a $0{+}1$D superconformal bath. From the thermodynamics of this regime we obtained a parameter-free relation between the Sommerfeld coefficient and the static spin and charge susceptibilities, a prediction by which the framework can be tested.

\paragraph*{Acknowledgements.}
I am grateful to Sergey Frolov and Niklas Beisert for introducing me to the electronic interpretation of $\mathfrak{su}(2|2)$, from which this work grew. Much of the thinking behind it took shape during my years at Trinity College Dublin, the Max Planck Institute for the Physics of Complex Systems (MPI-PKS) in Dresden, the University of Amsterdam, and LPTMS at Université Paris-Saclay/CNRS, and I thank colleagues at each.

\appendix

\section{The Hubbard algebra}
\label{app:Hubbard}

Here we provide a summary of the exceptionally centrally extended Lie superalgebra $\mathfrak{su}(2|2)$, along with its mapping to the Hubbard operators. The central extension is inherited from $D(2,1;\alpha)$, as detailed in \secref{sec:partialcontraction}. We adopt the conventions of Ref.~\cite{QuinnHidden2021}, in which the sign of $q^\dagger_{\sigma\bullet}$ is flipped relative to the conventions of Refs.~\cite{Quinn2018,QuinnErten}, which gives the algebra a more symmetric form.

The $\mathfrak{su}(2|2)$ algebra is generated by two bosonic $\mathfrak{su}(2)$ triplets, $\vec{s}$ and $\vec{\eta}$, and eight fermionic generators $q_{\sigma\nu}$ and $q^\dagger_{\sigma\nu}$, with indices $\sigma \in \{\uparrow, \downarrow\}$ and $\nu \in \{\bullet, \circ\}$. The two $\mathfrak{su}(2)$ algebras mutually commute, and writing $s^\pm = s^x \pm i s^y$ and $\eta^\pm = \eta^x \pm i \eta^y$, their $\mathfrak{su}(2)$ relations are given by
\begin{equation}
[s^+,s^-]=2s^z,\quad [s^z,s^\pm]=\pm s^\pm,
\end{equation}
\begin{equation}
[\eta^+,\eta^-]=2\eta^z,\quad [\eta^z,\eta^\pm]=\pm \eta^\pm
\end{equation}
The fermionic generators transform in the fundamental representations of the two $\mathfrak{su}(2)$:
\begin{equation}
[s^z, q^\dagger_{\sigma\nu}] = \frac{1}{2}\sigma q^\dagger_{\sigma\nu},\quad 
[s^\pm, q^\dagger_{\mp\nu}] = -q^\dagger_{\pm\nu},\quad
[s^\pm, q_{\pm\nu}] = q_{\mp\nu},
\end{equation}
\begin{equation}
[\eta^z, q^\dagger_{\sigma\nu}] = \frac{1}{2}\nu q^\dagger_{\sigma\nu},\quad 
[\eta^\pm, q^\dagger_{\sigma\mp}] = -q^\dagger_{\sigma\pm},\quad
[\eta^\pm, q_{\sigma\pm}] = q_{\sigma\mp},
\end{equation}
where as coefficients the indices take the values $+1$ for $\{\uparrow, \bullet\}$ and $-1$ for $\{\downarrow, \circ\}$. The anti-commutators of the fermionic generators close back onto the bosonic sector, parameterized by a continuous parameter $\kappa$:
\begin{equation}
\begin{aligned}
&\{ q_{\sigma\nu}, q^\dagger_{\sigma\nu}\} = \frac{1+\kappa^2}{4} - \kappa \left(\sigma s^z - \nu\eta^z\right), \\
&\{ q_{\downarrow\nu}, q^\dagger_{\uparrow\nu}\} = \kappa s^+, \qquad \{ q_{\sigma\circ}, q^\dagger_{\sigma\bullet}\} = -\kappa \eta^+, \\
&\{ q_{\uparrow\nu}, q^\dagger_{\downarrow\nu}\} = \kappa s^-, \qquad \{ q_{\sigma\bullet}, q^\dagger_{\sigma\circ}\} = -\kappa \eta^-, \\
&\{ q_{\sigma\nu}, q_{\sigma'\nu'}\} = \{ q^\dagger_{\sigma\nu}, q^\dagger_{\sigma'\nu'}\} = \frac{1-\kappa^2}{4}\epsilon_{\sigma\sigma'}\epsilon_{\nu\nu'},
\end{aligned}
\end{equation}
where as coefficients $\sigma$, $\nu$ take values $+1$ for $\{\uparrow, \bullet\}$ and $-1$ for $\{\downarrow, \circ\}$, and $\epsilon$ is the antisymmetric tensor with convention $\epsilon_{\downarrow\uparrow}=\epsilon_{\circ\bullet}=1$.
The algebra is extended to $\mathfrak{u}(2|2)$ through the additional generator $\theta$ obeying
\begin{equation}
\begin{aligned}
[\theta,q^\dagger_{\sigma\nu}] &= \frac{1+\kappa^2}{4\kappa} q^\dagger_{\sigma\nu} - \frac{1-\kappa^2}{4\kappa}\epsilon_{\sigma\sigma'}\epsilon_{\nu\nu'}q_{\sigma'\nu'} \\
[\theta,q_{\sigma\nu}] &= - \frac{1+\kappa^2}{4\kappa} q_{\sigma\nu} + \frac{1-\kappa^2}{4\kappa}\epsilon_{\sigma\sigma'}\epsilon_{\nu\nu'}q^\dagger_{\sigma'\nu'}.
\end{aligned}
\end{equation}

Precisely at $\kappa=1$, the algebra maps to the Hubbard projection operators, $X_{ab} = |a\rangle\langle b|$, satisfying the graded algebra:
\begin{equation}
[X_{ab}, X_{cd}\}_{\pm} = \delta_{bc} X_{ad} \pm \delta_{ad} X_{cb},
\end{equation}
where the anti-commutator is used if and only if both operators are fermionic, and the commutator is used otherwise. The explicit mapping is:
\begin{equation}
\begin{aligned}
X_{\nu\sigma} &= \bar{\nu} q^\dagger_{\bar{\sigma}\nu}, \quad & X_{\nu\nu} &= \nu \eta^z + \theta + \frac{1}{4}, \quad & X_{\bullet\circ} &= \eta^+, \quad & X_{\circ\bullet} &= \eta^-, \\
X_{\sigma\nu} &= \bar{\nu} q_{\bar{\sigma}\nu}, \quad & X_{\sigma\sigma} &= \sigma s^z - \theta + \frac{1}{4}, \quad & X_{\uparrow\downarrow} &= s^+, \quad & X_{\downarrow\uparrow} &= s^-,
\end{aligned}
\end{equation}
where $\bar{\sigma} = -\sigma$ and $\bar{\nu} = -\nu$.

We now express the $\mathfrak{su}(2|2)$ generators through the canonical fermion $c_\sigma$, $c^\dagger_\sigma$ and number operator $n_\sigma=c^\dagger_\sigma c_\sigma$. The fermionic generators are:
\begin{equation}
\begin{aligned} &q^\dagger_{\sigma\circ} = \frac{1+\kappa}{2} c_{\bar{\sigma}} - \kappa n_\sigma c_{\bar{\sigma}}, \qquad q^\dagger_{\sigma\bullet} = \sigma \big( \frac{1-\kappa}{2} c^\dagger_\sigma + \kappa n_{\bar{\sigma}} c^\dagger_\sigma \big), \\
&q_{\sigma\circ} = \frac{1+\kappa}{2} c^\dagger_{\bar{\sigma}} - \kappa n_\sigma c^\dagger_{\bar{\sigma}}, \qquad q_{\sigma\bullet} = \sigma \big( \frac{1-\kappa}{2} c_\sigma + \kappa n_{\bar{\sigma}} c_\sigma \big),
\end{aligned}
\end{equation}
and inverting for $c^\dagger_\sigma$ yields the splitting relations:
\begin{equation}
c^\dagger_{\downarrow} = q_{\uparrow\circ}-q^\dagger_{\downarrow\bullet}, \qquad c^\dagger_{\uparrow} = q_{\downarrow\circ}+q^\dagger_{\uparrow\bullet}.
\end{equation}
The spin and charge generators are:
\begin{equation}
\begin{aligned} 
&s^+ = c^\dagger_\uparrow c_\downarrow, \qquad s^- = c^\dagger_\downarrow c_\uparrow, \qquad s^z = \tfrac{1}{2}(n_\uparrow - n_\downarrow), \\
&\eta^+ = c^\dagger_\downarrow c^\dagger_\uparrow, \qquad \eta^- = c_\uparrow c_\downarrow, \qquad \eta^z = \tfrac{1}{2}(n_\uparrow + n_\downarrow - 1),
 \end{aligned}
\end{equation}
and together obey the Casimir identity $\vec{s}\cdot\vec{s} + \vec{\eta}\cdot\vec{\eta}=\frac{3}{4}$. The Hubbard generator is:
\begin{equation}
\theta =  \tfrac{1}{3}\big(\vec{\eta}\cdot\vec{\eta} - \vec{s}\cdot\vec{s}\big)
=\big(n_\downarrow - \tfrac{1}{2}\big)\big(n_\uparrow - \tfrac{1}{2}\big).
\end{equation}

\section{The bare Hubbard bands}
\label{app:bare_hubbard}

Here we present the bare Hubbard-mode Green's functions computed for the correlated hopping term, Eq.~\eqref{eq:T_qq}, of \secref{sec:modelham}, following Ref.~\cite{Quinn2018}. As outlined in \secref{sec:hubbardalg}, we retain only the scalar contributions to the anticommutation relations, Eq.~\eqref{eq:q_ac}, which fixes the bare modes exactly in $\kappa$. Upon computing the $q^\dagger_{\sigma\nu}$ matrix Green's function, the electronic Green's function is obtained via the splitting relations, Eq.~\eqref{eq:splitting}, resulting in (adapting for a factor of $\kappa$ in Ref.~\cite{Quinn2018}'s definition of $\theta$):
\begin{equation}\label{eq:Gelec}
G_p(\omega) =
\cfrac{1}{\;\omega+\mu-\cfrac{\varepsilon_p}{1+\kappa^2}
 - \cfrac{1}{4}\,
   \cfrac{\big(\kappa U+\lambda\varepsilon_p\big)^2}
         {\omega+\mu-\cfrac{\kappa^2\varepsilon_p}{1+\kappa^2}}\;} ,
\end{equation}
where $\varepsilon_p$ is the bare dispersion.
This resolves into two poles:
\begin{equation}\label{eq:Gsplit}
G_p(\omega) = \frac{a_{p\circ}}{\omega+\mu-\omega_{p\circ}}
            + \frac{a_{p\bullet}}{\omega+\mu-\omega_{p\bullet}},
\end{equation}
with $\nu = \circ, \bullet$ labelling the split bands ($\nu$ takes values $-1$ for $\circ$ and $+1$ for $\bullet$):
\begin{equation}\label{eq:bands}
\omega_{p\nu} = \frac{\varepsilon_p}{2}
 + \frac{\nu}{2}\sqrt{\left(\frac{1-\kappa^2}{1+\kappa^2}\right)^{\!2}\varepsilon_p^2
   +\big(\kappa U+\lambda\varepsilon_p\big)^2},
\end{equation}
which share the electronic spectral weight between them
\begin{equation}\label{eq:weights}
a_{p\nu} = \frac{1}{2}
 + \frac{\nu}{2}\,
   \frac{\dfrac{1-\kappa^2}{1+\kappa^2}\,\varepsilon_p}
        {\sqrt{\left(\dfrac{1-\kappa^2}{1+\kappa^2}\right)^{\!2}\varepsilon_p^2
         + \big(\kappa U+ \lambda \varepsilon_p\big)^2}}.
\end{equation}
In \secref{sec:discussion} we consider setting $t_\pm=0$ and $\lambda=0$, so that $\kappa=1$ via Eq.~\eqref{eq:CH_params}, resulting in Eq.~\eqref{eq:G_kappa_1}.

\section{The Kondo degree of freedom}
\label{app:kondo}

We call the combination of an electron and a local $\mathfrak{su}(2)$ spin a Kondo degree of freedom. We denote the spin as $\vec{S}$, obeying $\vec{S}\cdot\vec{S}=S(S+1)$, with $S$ the size of the spin. The Kondo degree of freedom then has a basis of $4\times (2S+1)$ states.

As for the electron, we consider two distinct formulations of the Kondo degree of freedom. The canonical formulation is the conventional one, with the canonical electron $\mathfrak{su}(1|1)\oplus\mathfrak{su}(1|1)$ and the $\mathfrak{su}(2)$ spin combining as $\mathfrak{su}(1|1)\oplus\mathfrak{su}(1|1)\oplus\mathfrak{su}(2)$. This underlies a description of a LFL interacting with local spins that are free to order.

The second formulation is a natural generalisation of the Hubbard formulation of the electron \cite{QuinnErten,Quinn2021}. Just as the non-canonical algebra $\mathfrak{su}(2)$ admits a family of representations, so too does $\mathfrak{su}(2|2)$. These can be directly interpreted as incorporating a spin $\vec{S}$, and have dimensionality $4 \times(2S+1)$. The Hubbard electron corresponds to the additional spin being a singlet $S=0$. The spin enters the algebra through the combination,
\begin{equation}
\vec{\Sigma} = \vec{s} + \vec{S},
\end{equation}
which obeys the spin $\mathfrak{su}(2)$ relations. This enters the fermionic anticommutation relations as follows:
\begin{equation}\label{eq:qS}
\begin{aligned}
&  \{ q_{\sigma\nu}, q^\dagger_{\sigma\nu}\} = \frac{1+\kappa^2}{4} - \frac{\kappa}{2S+1}\left(\sigma \Sigma^z  - \nu \eta^z\right),\\
& \{q_{\downarrow\nu}, q^\dagger_{\uparrow\nu}\} = \frac{\kappa}{2S+1} \Sigma^+, \qquad
\{ q_{\sigma\circ}, q^\dagger_{\sigma\bullet}\} = -\frac{\kappa}{2S+1} \eta^+,\\
& \{ q_{\uparrow\nu}, q^\dagger_{\downarrow\nu}\} = \frac{\kappa}{2S+1} \Sigma^-, \qquad
\{q_{\sigma\bullet}, q^\dagger_{\sigma\circ}\} = -\frac{\kappa}{2S+1} \eta^-,\\
&  \{ q_{\sigma\nu}, q_{\sigma'\nu'}\} =
\{ q^\dagger_{\sigma\nu}, q^\dagger_{\sigma'\nu'}\} = \frac{1-\kappa^2}{4}\epsilon_{\sigma \sigma'}\epsilon_{\nu \nu'}.
\end{aligned}
\end{equation} 
All other relations are the same as those for the Hubbard electron. The additional generator $\theta$ obeying, 
\begin{equation}
\begin{aligned} [\theta,q^\dagger_{\sigma\nu}] &= \frac{1+\kappa^2}{4\kappa} q^\dagger_{\sigma\nu} - \frac{1-\kappa^2}{4\kappa}\epsilon_{\sigma\sigma'}\epsilon_{\nu\nu'}q_{\sigma'\nu'} \\ [\theta,q_{\sigma\nu}] &= - \frac{1+\kappa^2}{4\kappa} q_{\sigma\nu} + \frac{1-\kappa^2}{4\kappa}\epsilon_{\sigma\sigma'}\epsilon_{\nu\nu'}q^\dagger_{\sigma'\nu'}, \end{aligned}
\end{equation}
can be expressed through the $\mathfrak{su}(2|2)$ generators as follows,
\begin{equation}\label{eqn:Theta}
\theta = \frac{1}{2} -\frac{1}{2S+1}\left(\vec{\Sigma} \cdot \vec{\Sigma} + \frac{1}{3}\vec{\eta}\cdot \vec{\eta}\right).
\end{equation}
Alternatively, employing $\vec{S}\cdot\vec{S}=S(S+1)$ and $\vec{s}\cdot\vec{s} + \vec{\eta}\cdot\vec{\eta}  = \frac{3}{4}$, we can relate it to the Kondo coupling:
\begin{equation}\label{eq:kondo}
\vec{s} \cdot \vec{S} = \frac{1}{3}\vec{\eta}\cdot\vec{\eta} - \frac{2S+1}{2} \theta - \frac{1+4S^2}{8}.
\end{equation}
As for the Hubbard electron, the fermionic Kondo generators again manifest a splitting of the electron:
\begin{equation}\label{eq:splitting_S}
c^\dagger_{\downarrow} = q_{\uparrow\circ}-q^\dagger_{\downarrow\bullet},\quad
c^\dagger_{\uparrow} = q_{\downarrow\circ}+q^\dagger_{\uparrow\bullet}.
\end{equation}
They can be expressed in terms of the canonical electron and spin as follows,
\begin{equation}\label{eqn:defqS}
\begin{aligned}
&q^\dagger_{\downarrow\circ} = \frac{1}{2}c_{\uparrow} + \frac{\kappa}{2S+1} \left( \frac{1}{2}c_{\uparrow} -n_{\downarrow}c_{\uparrow} + c_{\downarrow} S^- + c_{\uparrow} S^z\right),\\
&q^\dagger_{\uparrow\circ} =\frac{1}{2}c_{\downarrow} +  \frac{\kappa}{2S+1} \left( \frac{1}{2}c_{\downarrow} - n_{\uparrow}c_{\downarrow} + c_{\uparrow} S^+ - c_{\downarrow} S^z\right),\\
&q^\dagger_{\downarrow\bullet} = -\frac{1}{2}c^\dagger_{\downarrow} + \frac{\kappa}{2S+1} \left( \frac{1}{2}c^\dagger_{\downarrow}- n_{\uparrow}c^\dagger_{\downarrow} + c^\dagger_{\uparrow} S^- - c^\dagger_{\downarrow} S^z\right),\\
&q^\dagger_{\uparrow\bullet} =\frac{1}{2}c^\dagger_{\uparrow} - \frac{\kappa}{2S+1} \left( \frac{1}{2}c^\dagger_{\uparrow} -n_{\downarrow}c^\dagger_{\uparrow} + c^\dagger_{\downarrow} S^+ + c^\dagger_{\uparrow} S^z\right),
\end{aligned}
\end{equation} 
which satisfy Eq.~\eqref{eq:qS} upon use of the Casimir identity $\vec{S}\cdot\vec{S}=S(S+1)$.

This Hubbard formulation of the Kondo degree of freedom provides a HFL description of the Kondo lattice. Here the spin is entwined with the electron, and not free to order by itself. The Kondo coupling is conveniently treated in this approach via Eq.~\eqref{eq:kondo}, from which we see it contributes to the emergent chemical potential driving a splitting of the electronic band. Again the electronic Luttinger sum rule is generically violated for this metallic state, and the competition between the LFL+spin and HFL offers a natural interpretation of the phase diagram of Heavy Fermion materials \cite{Stewart2001}, with the Heavy Fermi liquid identified as the HFL.

To draw the connection to the strange metal, let us highlight the contractions of $D(2,1;\alpha)$ of \secref{sec:algebra} in this setting. For the maximal contraction, the spin belongs to the decoupled part of the representation. For the partial contraction, the shortening condition, Eq.~\eqref{eq:shortening}, generalises to: 
\begin{equation}
\tilde{D}^2 - |\tilde{E}|^2 = \frac{2S+1}{4},
\end{equation}
which yields the $4(2S+1)$-dimensional representations of the spin-$S$ Kondo degree of freedom \cite{Beisert2007}.

\section{The Exceptional Lie Superalgebra $D(2,1;\alpha)$}
\label{app:exceptional}

In \secref{sec:algebra} we considered a non-compact real form of the exceptional Lie superalgebra $D(2,1;\alpha)$. Here we detail the compact form, as it may help orient a reader unfamiliar with these structures. We also highlight some notable limits.

The algebra $D(2,1;\alpha)$ comprises three bosonic $\mathfrak{su}(2)$ triplets $J^i_j$, $J^a_b$, $J^\alpha_\beta$, and a fermionic octet $Q^{i a\alpha}$, with all indices taking values in $\{1,2\}$. The $\mathfrak{su}(2)$ factors commute with one another and obey the $\mathfrak{su}(2)$ relations internally:
\begin{equation}
\begin{aligned}
[J^i_j, J^k_l] &= \delta^k_j J^i_l - \delta^i_l J^k_j,\\
[J^a_b, J^c_d] &= \delta^c_b J^a_d - \delta^a_d J^c_b,\\
[J^\alpha_\beta, J^\gamma_\delta] &= \delta^\gamma_\beta J^\alpha_\delta - \delta^\alpha_\delta J^\gamma_\beta.
\end{aligned}
\end{equation}
The fermionic generators transform in the fundamental representation of each $\mathfrak{su}(2)$ factor:
\begin{equation}\begin{aligned}
[J^i_j, Q^{k a\alpha}] &= \delta^k_j Q^{i a\alpha} - \frac{1}{2}\delta^i_j Q^{k a\alpha},\\
[J^a_b, Q^{i c\alpha}] &= \delta^c_b Q^{i a\alpha} - \frac{1}{2}\delta^a_b Q^{i c\alpha},\\
[J^\alpha_\beta, Q^{i a\gamma}] &= \delta^\gamma_\beta Q^{i a\alpha} - \frac{1}{2}\delta^\alpha_\beta Q^{i a\gamma},
\end{aligned}\end{equation}
and their anticommutators close onto the bosonic generators:
 \begin{equation}
\{Q^{i a \alpha}, Q^{j b \beta}\} = c_1 \epsilon^{ab} \epsilon^{\alpha\beta} \epsilon^{ik} J^j_k + c_2 \epsilon^{ij} \epsilon^{\alpha\beta} \epsilon^{ac} J^b_c + c_3 \epsilon^{ij} \epsilon^{ab} \epsilon^{\alpha\gamma} J^\beta_\gamma,
\end{equation}
weighted by coefficients $c_1$, $c_2$, $c_3$ constrained to obey,
\begin{equation}
c_1 + c_2 + c_3 = 0,
\end{equation}
as the requirement of the super-Jacobi identity, and we set $\alpha=c_3/c_2$.

The non-compact real form employed in \secref{sec:algebra} is obtained by casting the $\mathfrak{su}(2)$ factor $J^i_j$ as a conformal $\mathfrak{sl}(2,\mathbb{R})$:
\begin{equation}
D = J^2_2 = -J^1_1,\quad E_+ =J^2_1,\quad E_- = J^1_2,
\end{equation}
via the reality conditions $D^\dagger = D$ and $E_\pm^\dagger = - E_\mp$. The other two $\mathfrak{su}(2)$ remain compact, $(J^a_b)^\dagger = J^b_a$ and $(J^\alpha_\beta)^\dagger = J^\beta_\alpha$, and the mapping is completed by splitting the fermionic octet into quartets of positive and negative conformal weight, $(Q^{a\alpha})^\dagger = Q^\dagger_{a\alpha}$:
\begin{equation}
Q^{a\alpha} = Q^{1 a\alpha},\quad Q^\dagger_{a\alpha} = \epsilon_{ab} \epsilon_{\alpha\beta} Q^{2 b\beta}.
\end{equation}

There are some notable limits of $\alpha$ which we employ. When $\alpha\to1$ the algebra becomes $\mathfrak{osp}(4|2)$: the two $\mathfrak{su}(2)$ factors are on a par and combine as $\mathfrak{so}(4) \simeq \mathfrak{su}(2) \oplus \mathfrak{su}(2)$, and the conformal $\mathfrak{sl}(2, \mathbb{R})$ provides the $\mathfrak{sp}(2)$. When $\alpha\to0$ or $\alpha\to\infty$, the algebra becomes $\mathfrak{psu}(1,1|2)$, upon decoupling one of the $\mathfrak{su}(2)$ factors and rescaling the generators. The projective $\mathfrak{p}$ denotes the absence of the central element. We consider the limit $\alpha\to-1$ as a contraction of the conformal sector in \secref{sec:partialcontraction}.

\section{Super-Virasoro extension of $D(2,1;\alpha)$}
\label{app:vira}

Here we detail the linearised super-Virasoro extension of $D(2,1;\alpha)$, known as the large $\mathcal{N}=4$ superconformal algebra.

The algebra governs the local dynamics of the $0+1$D bath studied in \secref{sec:eosm}. Its generators correspond to the Fourier modes of the fluctuating currents on the thermal circle. We write indices $m,n \in \mathbb{Z}$ for bosonic modes, $r,s \in \mathbb{Z} + 1/2$ for fermionic modes. We employ the doublet notation of \secref{sec:algebra} for the generators, with the electronic identification given by Eq.~\eqref{eq:electron_id}. The conformal sector gets lifted to a Virasoro algebra $L_n$, with $L_0=D$, $L_1=E_-$,  $L_{-1}=-E_+$. We denote the auxiliary $U(1)$ current as $Y_n$, and its associated Majorana fermions as $\Psi^{a\alpha}_r$, with reality condition $\Psi^\dagger_{a\alpha,r} = \epsilon_{ab}\epsilon_{\alpha\beta}\Psi^{b\beta}_r$.

\begin{itemize}
\item Conformal sector:
\begin{equation}\begin{split}
[L_m, L_n] &= (m-n)L_{m+n} + \frac{c}{12}m(m^2-1)\delta_{m+n,0}, \qquad c = \frac{6k_sk_c}{k_s+k_c} \\
[L_m, J^a_{b,n}] &= -n\,J^a_{b,m+n}, \quad [L_m, J^\alpha_{\beta,n}] = -n\,J^\alpha_{\beta,m+n}, \quad [L_m, Y_n] = -n\,Y_{m+n} \\
[L_m, Q^{a\alpha}_r] &= \left(\frac{m}{2}-r\right)Q^{a\alpha}_{m+r}, \qquad [L_m, \Psi^{a\alpha}_r] = -\left(\frac{m}{2}+r\right)\Psi^{a\alpha}_{m+r}
\end{split}\end{equation}

\item Bosonic currents:
\begin{equation}\begin{split}
[J^a_{b,m}, J^c_{d,n}] &= \delta^c_b J^a_{d,m+n} - \delta^a_d J^c_{b,m+n} +k_s \,m \,\delta_{m+n,0}\Big(\delta^c_b\delta^a_d - \frac{1}{2}\delta^a_b\delta^c_d\Big) \\
[J^\alpha_{\beta,m}, J^\gamma_{\delta,n}] &= \delta^\gamma_\beta J^\alpha_{\delta,m+n} - \delta^\alpha_\delta J^\gamma_{\beta,m+n} + k_c \,m\,\delta_{m+n,0}\Big(\delta^\gamma_\beta\delta^\alpha_\delta - \frac{1}{2}\delta^\alpha_\beta\delta^\gamma_\delta\Big) \\
[J^a_{b,m}, J^\alpha_{\beta,n}] &= 0 \\
[Y_m, J^a_{b,n}] &= [Y_m, J^\alpha_{\beta,n}] = 0 \\
[Y_m, Y_n] &= \frac{2}{c_1^2} (k_s+k_c)\,m\,\delta_{m+n,0}
\end{split}\end{equation}

\item Current action on fermions:
\begin{equation}\begin{split}
[J^a_{b,m}, Q^{c\gamma}_r] &= \Big(\delta^c_b Q^{a\gamma}_{m+r} - \frac{1}{2}\delta^a_b Q^{c\gamma}_{m+r}\Big) - c_1 m\frac{k_s}{k_s+k_c} \big(\delta^c_b\Psi^{a\gamma}_{m+r} - \frac{1}{2}\delta^a_b\Psi^{c\gamma}_{m+r}\big) \\
[J^\alpha_{\beta,m}, Q^{c\gamma}_r] &= \big(\delta^\gamma_\beta Q^{c\alpha}_{m+r} - \frac{1}{2}\delta^\alpha_\beta Q^{c\gamma}_{m+r}\big) + c_1 m\frac{k_c}{k_s+k_c} \big(\delta^\gamma_\beta\Psi^{c\alpha}_{m+r} - \frac{1}{2}\delta^\alpha_\beta\Psi^{c\gamma}_{m+r}\big) \\
[J^a_{b,m}, \Psi^{c\gamma}_r] &= \delta^c_b\Psi^{a\gamma}_{m+r} - \frac{1}{2}\delta^a_b\Psi^{c\gamma}_{m+r} \\
[J^\alpha_{\beta,m}, \Psi^{c\gamma}_r] &= \delta^\gamma_\beta\Psi^{c\alpha}_{m+r} - \frac{1}{2}\delta^\alpha_\beta\Psi^{c\gamma}_{m+r}
\end{split}\end{equation}

\item Linearising couplings:
\begin{equation}\begin{split}
[Y_m, Q^{a\alpha}_r] &= m\,\Psi^{a\alpha}_{m+r} \\
[Y_m, \Psi^{a\alpha}_r] &= 0 \\
\{\Psi^{a\alpha}_r, Q^\dagger_{b\beta,s}\} &= \delta^\alpha_\beta J^a_{b,r+s} - \delta^a_b J^\alpha_{\beta,r+s} + \frac{c_1}{2}\delta^a_b\delta^\alpha_\beta\,Y_{r+s}
\end{split}\end{equation}

\item Fermionic closures:
\begin{align}
\{Q^{a\alpha}_r, Q^\dagger_{b\beta,s}\} &= c_1\,\delta^a_b\delta^\alpha_\beta\Big(L_{r+s} + \frac{c}{6}\big(r^2-\frac{1}{4}\big)\delta_{r+s,0}\Big) - (r-s)\big(c_2\,\delta^\alpha_\beta J^a_{b,r+s} + c_3\,\delta^a_b J^\alpha_{\beta,r+s}\big) \\
\{Q^{a\alpha}_r, Q^{b\beta}_s\} &= c_1\,\epsilon^{ab}\epsilon^{\alpha\beta}\Big(L_{r+s} + \frac{c}{6}\big(r^2-\frac{1}{4}\big)\delta_{r+s,0}\Big) - (r-s)\big(c_2\,\epsilon^{\alpha\beta}\epsilon^{bc} J^a_{c,r+s} + c_3\,\epsilon^{ab}\epsilon^{\beta\gamma} J^\alpha_{\gamma,r+s}\big) \notag
\end{align}
\begin{equation}\begin{split}
\{\Psi^{a\alpha}_r, \Psi^\dagger_{b\beta,s}\} &= \frac{1}{c_1}(k_s+k_c)\,\delta^a_b\delta^\alpha_\beta\,\delta_{r+s,0} \\
\{\Psi^{a\alpha}_r, \Psi^{b\beta}_s\} &= \frac{1}{c_1}(k_s+k_c)\,\epsilon^{ab}\epsilon^{\alpha\beta}\,\delta_{r+s,0}
\end{split}\end{equation}

\item Structure constants, constrained by the super-Jacobi identity:
\begin{equation}
c_1+c_2+c_3=0,\quad
c_2 = -c_1\frac{k_c}{k_s+k_c},\quad c_3 = -c_1\frac{k_s}{k_s+k_c},\quad \alpha = \frac{c_3}{c_2} = \frac{k_s}{k_c}.
\end{equation}

\end{itemize}

\subsection*{Super-Virasoro extension of $PSU(1,1|2)$}

The super-Virasoro extension of $PSU(1,1|2)$ is the small $\mathcal{N}=4$ superconformal algebra, and within our framework this is the superconformal algebra of Mott criticality, \secref{sec:mottcriticality}. This is obtained as the limit $k_c\to\infty$ of the above, in which $J^\alpha_{\beta,n}$, $Y_n$ and $\Psi^{b\beta}_r$ decouple, the structure constants reduce to $c_2=-c_1$, $c_3=0$, and the central charge becomes $c=6 k_s$.

\section{One-loop correction}
\label{app:one-loop}

The exact source-dependent partition function of the large $\mathcal N=4$ Schwarzian theory was computed by Heydeman, Shi and Turiaci \cite{HeydemanShiTuriaci2025}. Mapping this to the electronic setting (\secref{sec:electronicinterp}), it takes the form\footnote{We set the source for the decoupled
$U(1)$ charge $Y_0$ to zero, as it does not contribute to the physical spin and charge responses we consider. The residual sum
over $U(1)$ winding sectors gives a correction exponentially small in $MT$, and is dropped.}:
\begin{equation}
\label{eq:app_Z}
Z(\beta, h_s, h_c)
= e^{S_0+2\pi^2 M/\beta}
\sum_{m,n\in\mathbb{Z}}
Z^{(m,n)}_{\text{1-loop}}\,
\exp\!\left[\frac{\lambda_s M}{2\beta}(\beta h_s + 4\pi i m)^2
+ \frac{\lambda_c M}{2\beta}(\beta h_c + 4\pi i n)^2\right],
\end{equation}
with 
\begin{equation}
\label{eq:app_1loop}
\begin{split}
Z^{(m,n)}_{\text{1-loop}}
=\frac{M}{\beta}\sqrt{\frac{2\pi(1+\alpha)^2}{\alpha}}&\,
\underbrace{
\frac{
    \cosh\frac{\beta h_s+\beta h_c}{4}\,\cosh\frac{\beta h_s-\beta h_c}{4}}
{\big[1+4(\frac{\beta h_s+\beta h_c}{4\pi}+i(m{+}n))^2\big]
 \big[1+4(\frac{\beta h_s-\beta h_c}{4\pi}+i(m{-}n))^2\big]}
}_{\text{from fermion partners of reparam.\ mode}}\\[1.2em]
&\,\times\;
\underbrace{
\frac{
    \big(\frac{\beta h_s}{4\pi}+im\big)
    \big(\frac{\beta h_c}{4\pi}+in\big)
 }{\sinh\frac{\beta h_s}{2} \sinh\frac{\beta h_c}{2}}
}_{\text{from the } SU(2)\text{ current modes}}
\;\times\;
\underbrace{
4\cosh\tfrac{\beta h_s+\beta h_c}{4}\,\cosh\tfrac{\beta h_s-\beta h_c}{4}
}_{\text{from fermion partners of }U(1)\text{ mode}}.
\end{split}
\end{equation}

Within the conformal
window, $M\beta\gg1$, the dominant contribution comes from the zero-winding sector,
\begin{equation}
\label{eq:app_logZ0}
\log Z \approx S_0 + \frac{2\pi^2 M}{\beta}
+ \frac{M\beta}{2}\big(\lambda_s h_s^2 + \lambda_c h_c^2\big)
+ \log Z^{(0,0)}_{\text{1-loop}}.
\end{equation}
Up to the one-loop correction, this reproduces the semiclassical saddle-point analysis of \secref{sec:thermo}. The one-loop corrections to the entropy and specific heat, Eqs.~\eqref{eq:entropy_oneloop}-\eqref{eq:Cv_oneloop} with $p=1$, follow immediately upon setting $h_s=h_c=0$. For the one-loop correction to the static susceptibilities, we focus on spin, 
setting $h_c=0$ and computing:
\begin{equation}
\begin{split}
\frac{\partial^2}{\partial(\beta h_s)^2}\log Z^{(0,0)}_{\text{1-loop}}\bigg|_{0}
&= \frac{\partial^2}{\partial(\beta h_s)^2}
\Big[
\log\cosh^4\tfrac{\beta h_s}{4}
+ \log\frac{\beta h_s/4\pi}{\sinh\frac{\beta h_s}{2}}
- \log\!\Big(1+\tfrac{(\beta h_s)^2}{4\pi^2}\Big)^{2}
\Big]_{0}
\\
&=  \frac16 - \frac{1}{\pi^2},
\end{split}
\end{equation}
and charge follows identically. We thus obtain,
\begin{equation}
\label{eq:app_chi}
\chi_s = (1+\alpha)M + \Big(\frac16-\frac{1}{\pi^2}\Big)\frac{1}{T},
\qquad
\chi_c = \frac{1+\alpha}{\alpha}M + \Big(\frac16-\frac{1}{\pi^2}\Big)\frac{1}{T},
\end{equation}
up to $O(e^{-8\pi^2\lambda_{s,c}MT})$ winding corrections. We also note that the mixed susceptibility, $\chi_{sc} = \frac{1}{\beta} \frac{\partial^2 \log Z}{\partial  h_s \partial h_c}\big|_0$, vanishes at this order.

\bibliographystyle{physref}
\bibliography{refs}

\end{document}